\documentclass[aps,
               prc,
               twocolumn,
               superscriptaddress,
               floatfix,
               amsmath,
               amssymb,
               nofootinbib,
               longbibliography,
               reqno,
               tbtags]{revtex4-2}

\usepackage[utf8]{inputenc}
\usepackage[T1]{fontenc}
\usepackage{soul}
\usepackage[dvipsnames]{xcolor}
\usepackage{graphicx}
\usepackage[ocgcolorlinks,citecolor=blue,linkcolor=blue,urlcolor=blue]{hyperref}
\usepackage{scalefnt}
\usepackage{xspace}
\usepackage{times,txfonts}
\usepackage{pifont}
\usepackage{bm}
\usepackage{soul}

\usepackage{tabularx}
\usepackage{booktabs}
\usepackage{physics}
\usepackage[capitalise]{cleveref}

\renewcommand{\v}[1]{\mathbf{#1}}
\newcommand{\eg}{e.g.}

\newcommand{\clebsch}[6]{\mathcal{C}_{#1 #2 #3 #4}^{#5 #6} \,}

\newcommand{\ntwolo}{N$^2$LO\xspace}
\newcommand{\nthreelo}{N$^3$LO\xspace}

\newcommand{\sixj}[6]{
     \left\{ \begin{array}{ccc}
              #1 & #2 & #3 \\
              #4 & #5 & #6
            \end{array}  \right\} }
\newcommand{\ninej}[9]{
     \left\{ \begin{array}{ccc}
              #1 & #2 & #3 \\
              #4 & #5 & #6 \\
              #7 & #8 & #9
            \end{array}  \right\} }

\begin{document}

\title{Magnetic dipole operator from chiral effective field theory for many-body expansion methods}

\author{R.~Seutin}
\email{rseutin@theorie.ikp.physik.tu-darmstadt.de}
\affiliation{Max-Planck-Institut f\"ur Kernphysik, Saupfercheckweg 1, 69117 Heidelberg, Germany}
\affiliation{Technische Universit\"at Darmstadt, Department of Physics, 64289 Darmstadt, Germany}
\affiliation{ExtreMe Matter Institute EMMI, GSI Helmholtzzentrum f\"ur Schwerionenforschung GmbH, 64291 Darmstadt, Germany}

\author{O.~J.~Hernandez}
\affiliation{Institut f\"ur Kernphysik and PRISMA Cluster of Excellence, Johannes Gutenberg-Universit\"at Mainz, 55128 Mainz, Germany}

\author{T.~Miyagi}
\email{miyagi@theorie.ikp.physik.tu-darmstadt.de}
\affiliation{Technische Universit\"at Darmstadt, Department of Physics, 64289 Darmstadt, Germany}
\affiliation{ExtreMe Matter Institute EMMI, GSI Helmholtzzentrum f\"ur Schwerionenforschung GmbH, 64291 Darmstadt, Germany}
\affiliation{Max-Planck-Institut f\"ur Kernphysik, Saupfercheckweg 1, 69117 Heidelberg, Germany}

\author{S.~Bacca}
\email{s.bacca@uni-mainz.de}
\affiliation{Institut f\"ur Kernphysik and PRISMA Cluster of Excellence, Johannes Gutenberg-Universit\"at Mainz, 55128 Mainz, Germany}
\affiliation{Helmholtz Institute Mainz, GSI Helmholtzzentrum f\"ur Schwerionenforschung GmbH, 64289 Darmstadt, Germany}

\author{K.~Hebeler}
\email{kai.hebeler@physik.tu-darmstadt.de}
\affiliation{Technische Universit\"at Darmstadt, Department of Physics, 64289 Darmstadt, Germany}
\affiliation{ExtreMe Matter Institute EMMI, GSI Helmholtzzentrum f\"ur Schwerionenforschung GmbH, 64291 Darmstadt, Germany}
\affiliation{Max-Planck-Institut f\"ur Kernphysik, Saupfercheckweg 1, 69117 Heidelberg, Germany}

\author{S.~K\"onig}
\email{skoenig@ncsu.edu}
\affiliation{North Carolina State University, Department of Physics, Raleigh, NC 27695, USA}
\affiliation{Technische Universit\"at Darmstadt, Department of Physics, 64289 Darmstadt, Germany}
\affiliation{ExtreMe Matter Institute EMMI, GSI Helmholtzzentrum f\"ur Schwerionenforschung GmbH, 64291 Darmstadt, Germany}

\author{A.~Schwenk}
\email{schwenk@physik.tu-darmstadt.de}
\affiliation{Technische Universit\"at Darmstadt, Department of Physics, 64289 Darmstadt, Germany}
\affiliation{ExtreMe Matter Institute EMMI, GSI Helmholtzzentrum f\"ur Schwerionenforschung GmbH, 64291 Darmstadt, Germany}
\affiliation{Max-Planck-Institut f\"ur Kernphysik, Saupfercheckweg 1, 69117 Heidelberg, Germany}

\begin{abstract}
Many-body approaches for atomic nuclei generally rely on a basis expansion of the nuclear states, interactions, and current operators. In this work, we derive the representation of the magnetic dipole operator in plane-wave and harmonic-oscillator basis states, as needed for Faddeev calculations of few-body systems or many-body calculations within, \eg, the no-core shell model, the in-medium similarity renormalization group, coupled-cluster theory, or the nuclear shell model. We focus in particular on the next-to-leading-order two-body contributions derived from chiral effective field theory. We provide detailed benchmarks and also comparisons with quantum Monte Carlo results for three-body systems. The derived operator matrix elements represent the basic input for studying magnetic properties of atomic nuclei based on chiral effective field theory.
\end{abstract}

\maketitle

\section{Introduction}

Calculating the electromagnetic structure of nuclei is a powerful tool to explore and test nuclear theory.
The weak electromagnetic coupling compared to the strong interaction allows for a perturbative treatment of these processes, so that the nuclear structure content can be separated with great control.
The electromagnetic interaction between the nucleus and external photons can in general be described by a current-current interaction. While quantum electrodynamics (QED) describes the current of the external probe, nuclear theory deals with the nuclear current. To first approximation, the interaction between the photon and an atomic nucleus can be expressed in terms of the sum of photon interactions with all the individual nucleons.  This approximation is equivalent to retaining only one-body contributions in the nuclear current, while all possible higher-body operators are neglected. Even though these leading terms provide the dominant contributions, higher-order contributions, especially from two-body operators are crucial for precise predictions of electromagnetic observables.

The modern approach to quantitatively understanding low-energy nuclear physics in terms of \emph{ab initio} calculations is based on effective field theory (EFT), most notably chiral EFT.
It provides a systematic expansion of the strong interaction between nucleons as well as electroweak interactions with a direct connection to the fundamental theory of quantum chromodynamics (QCD) and its symmetries~\cite{epelbaum2009,entemmachleidt2011,hammer2020}.
A power-counting scheme orders the expansion terms according to decreasing importance in powers of $(Q/\Lambda_\text{b})^\nu$, with $Q$ the typical momentum scale governing processes in the nucleus, which is of the order of the pion mass $m_\pi$, and $\Lambda_\text{b}$ the breakdown scale $\Lambda_{\rm b} = 500-600$\,MeV.
Leading order (LO) terms, i.e., $\nu=-2$ for electromagnetic currents, include the dominant one-body contributions mentioned earlier, while next-to-leading order (NLO) and next-to-next-to-leading order (N$^2$LO) terms, etc., add contributions of decreasing importance.
The systematic expansion provides a way to improve calculations and to determine uncertainties arising from neglected higher orders~\cite{epelbaum2015,furnstahl2015a}.
Furthermore, EFT provides a consistent derivation of nuclear forces and currents.
To date, there have been several efforts to derive electromagnetic nuclear currents within the framework of chiral EFT.
In Refs.~\cite{pastore2008,pastore2009,pastore2011} time-ordered perturbation theory was used to obtain current operator expressions up to next-to-next-to-next-to-leading order (N$^3$LO) in the chiral expansion, while Refs.~\cite{koelling2009,koelling2011,krebs2020} used the method of unitary transformation.
Both methods agree on the current operators at the order we employ in this work.
However, at higher orders disagreements occur; for a detailed discussion see Ref.~\cite{krebs2020}.

Calculating the electromagnetic structure of nuclei involves evaluating the electromagnetic nuclear current operator $J_\mu=(\rho, \v{j})$, with charge operator $\rho$ and three-vector current operator $\v{j}$, between initial and final states of the nuclear system $|i\rangle$ and $|f\rangle$.
The Fourier transform of the current operator contains information about the charge and magnetization densities inside the nucleus.
Because the nuclear states have a definite angular momentum, it is useful to decompose the nuclear current into its multipole components.
For example, the current operator $\v{j}$ can be expressed in terms of electric and magnetic multipole operators, the long-wavelength limits of which correspond to the electric and magnetic moment operators, where the magnetic dipole contribution is the focus of this work.
With the magnetic dipole operator, one can calculate ground-state properties like the nuclear magnetic moment, defined by
\begin{equation}
  \mu \equiv \mel{\xi J M=J\,}{\,\mu_{z}\,}{\,\xi J M=J},
\end{equation}
where $J$ and $M$ are the nuclear spin and its projection, respectively, and $\xi$ represents all other quantum numbers relevant to describe the state.
In addition, one can calculate magnetic transitions between nuclear states.
The probability of an initial state of the nucleus to emit or absorb a photon and transition to a final state is given by Fermi's golden rule~\cite{Suhonen2007}:
\begin{equation}
  \Gamma_{\gamma, i \rightarrow f} = \frac{2 \, (\lambda + 1)}{\lambda \big[ (2\lambda + 1)!! \big]^2} \, E_\gamma^{2\lambda+1} B(M \lambda; \xi_i J_i \rightarrow \xi_f J_f),
\end{equation}
where $\lambda$ is the angular momentum of the photon with energy $E_\gamma$.
We use units with $\hbar=c=\varepsilon_{0}=1$, and the last term above is the transition strength
\begin{equation}\label{eq:transition_strength}
  B(M \lambda; \xi J_i \rightarrow \xi J_f) \equiv \frac{1}{2J_i + 1} \big|\langle \xi_f J_f || \, O^\text{mag}_{\lambda} \, || \xi_i J_i \rangle\big|^2.
\end{equation}
Here, $O^\text{mag}_{\lambda}$ represents the magnetic multipole operator in the long-wavelength limit (momentum transfer $Q\to0$)~\cite{walecka1966, walecka1975}, with the reduced matrix element $\langle \xi_f J_f || \, O^\text{mag}_{\lambda} \, || \xi_i J_i \rangle$.
In the case $\lambda = 1$ the multipole operator is the magnetic dipole operator $O^\text{mag}_{1z} = \sqrt{\frac{3}{4\pi}}\mu_{z}$, and the corresponding transition is referred to as the dipole transition or $M1$ transition.

Studies of electromagnetic properties calculated with chiral EFT currents combined with nuclear states obtained from chiral EFT interactions or phenomenological potentials have to date been focused on few-nucleon systems and light nuclei.
Deuteron and trinucleon electromagnetic form factors, radii, and moments have been studied up to N$^3$LO in Refs.~\cite{phillips2000,meissner2001,phillips2003,phillips2007,phillips2008,epelbaum2012,piarulli2013,gasparyan2014,piarulli2013,NevoDinur2019,schiavilla2019}, with the most recent result for the charge and quadrupole form factors of the deuteron pushing the calculation to fifth order in the chiral expansion~\cite{filin2020,filin2021}.
In Ref.~\cite{pastore2013}, magnetic moments and electromagnetic transitions of light nuclei up to $A \leq 9$ have been calculated with a hybrid method, combining phenomenological wave functions with chiral magnetic dipole operators up to N$^3$LO, based on quantum Monte Carlo methods to solve the many-body problem.
More recently, the first full chiral EFT calculation of the ground-state magnetic moment and the lowest magnetic transition in $^6$Li has been presented~\cite{gayer2020}.
All studies identified that current operator contributions beyond LO are important to improve agreement with experimental magnetic properties, with two-body contributions entering at NLO having the largest impact.

\begin{figure*}[t!]
  \centering
  \includegraphics[width=\textwidth]{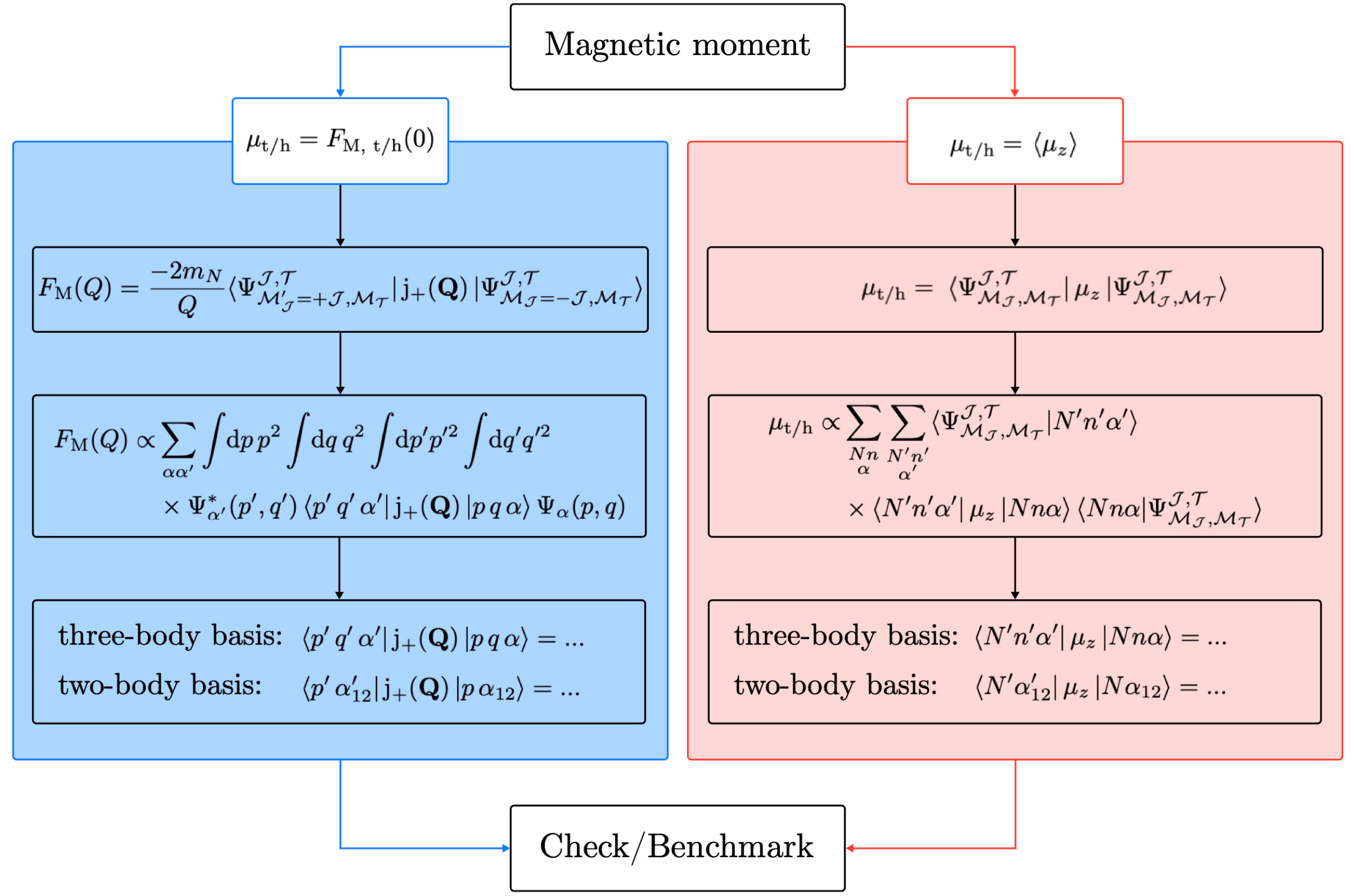}
  \caption{Schematic of the two methods used in this work to obtain the magnetic moments of the triton (t) and helion (h). The left part of the figure shows the steps for the current operator evaluated in a momentum-space basis to calculate the magnetic form factor, while the right part demonstrates the equivalent steps for the magnetic dipole operator evaluated between harmonic-oscillator basis states.}\label{fig:flowchart_paper}
\end{figure*}

Higher-order corrections to the current operator are clearly necessary for improving the agreement with experimental results and can provide an explanation to long-standing discrepancies between theory and experiment.
For example, the systematically smaller beta-decay rate in nuclei compared to free neutrons can, in part, be explained by the coupling of the weak force to two nucleons~\cite{Gysbers:2019uyb}.
In spite of this evidence for weak processes, the magnetic structure of heavier nuclei has so far only been studied without two-body currents (2BCs).
Most \emph{ab initio} many-body methods that calculate medium-mass nuclei are based upon basis expansion methods~\cite{Hagen2014,Hergert2016,Barb17SCGFlectnote,Stroberg2019,Tichai2020review}.
To perform computations, these frameworks require operators expanded in a computational basis that is commonly constructed based on harmonic-oscillator (HO) states.
In this work, we provide partial-wave matrix elements for the LO one-body-current and NLO two-body current operators in a two-body momentum-space basis as well as partial-wave matrix elements for the corresponding LO and NLO magnetic dipole operators in HO bases.
A straightforward implementation of these matrix elements can be used to calculate magnetic properties of medium-mass nuclei.
We validate our expressions by comparing the trinucleon magnetic moments obtained from the the magnetic form factors with Faddeev calculations against the magnetic dipole operator used in Jacobi no-core shell-model (NCSM) calculations.
\Cref{fig:flowchart_paper} displays the strategy of our trinucleon magnetic moments calculations.

This paper is organized as follows. In Sec.~\ref{sec:nuclear_magnetic_moments} we introduce the current operators that are employed and show how to obtain the magnetic dipole operators from them.
\Cref{sec:operator_matrix_elements} provides the expressions for the matrix elements of the various operators with respect to the different bases.
The results and comparison of the trinucleon magnetic moments are presented in Sec.~\ref{sec:results}.
Finally, we conclude in Sec.~\ref{sec:summary}.

\section{Nuclear magnetic moments}
\label{sec:nuclear_magnetic_moments}

Nuclear magnetic moments can be calculated using two related methods. The first uses the magnetic form factor at zero momentum transfer.
The form factor is the Fourier transform of the magnetization density of the nucleus and is obtained by calculating the expectation value of the nuclear current operator.
The second method obtains the nuclear magnetic moment by directly evaluating the magnetic moment operator, which is the long-wavelength limit of the dipole term of the multipole expansion of the current operator.
Below we specify the current and magnetic moment operators.

\subsection{Magnetic form factor normalization}

The one-body current operator at LO ($\sim e Q^{-2}$) is given in momentum space by~\cite{piarulli2013}
\begin{align}
  \v{j}^{(-2)}(\v{Q}) = \frac{e}{2m_N} \bigg( 2 \, e_{N}\big(Q^2\big) \, \v{K} + \mathrm{i} \, \mu_{N}\big(Q^2\big) \, \boldsymbol{\sigma} \times \v{Q} \bigg),
\label{eq:lo_current_operator}
\end{align}
where $e$ is the elementary charge, $m_N$ the nucleon mass, $\v{k}$ and $\v{k}'$ are the initial and final nucleon momenta, $\v{K} = (\v{k} + \v{k}')/2$, $\boldsymbol{\sigma}$ is the vector of Pauli spin matrices, and $\v{Q}$ is the spatial part of the momentum transfer associated with the photon $Q^2$ is $Q^\mu Q_\mu$. Momentum conservation requires that the relation $\v{k}' = \v{k} +\v{Q}$ holds. The functions $e_{N}\big(Q^2\big)$ and $\mu_{N}\big(Q^2\big)$ are given by
\begin{align}
  e_{N}\big(Q^2\big) &= \frac{G_\textnormal{E}^S\big(Q^2\big) + G_\textnormal{E}^V\big(Q^2\big) \tau_{z}}{2}, \\
  \mu_{N}\big(Q^2\big) &= \frac{G_\textnormal{M}^S\big(Q^2\big) + G_\textnormal{M}^V\big(Q^2\big) \tau_{z}}{2},
\end{align}
with $G_\textnormal{E}^{S/V}$ ($G_\textnormal{M}^{S/V}$) the isoscalar $S$ and isovector $V$ nucleon electric (magnetic) form factors, respectively.
At zero momentum transfer, the form factors are known to be $G_\textnormal{E}^S(0) = G_\textnormal{E}^V(0) = 1$, $G_\textnormal{M}^S(0) = 0.880$ $\mu_N$, and $G_\textnormal{M}^V(0) = 4.706$ $\mu_N$, where $\mu_{N} = e\hbar / 2m_{\rm proton}$ is the nuclear magneton. For all form-factor calculations in this work we employ the nucleon parametrization derived by Ye~\textit{et al.}~\cite{YE20188}. This parametrization includes two-photon exchange corrections as well as information from new high-precision electron-nucleon scattering data, including uncertainties.

\begin{figure}[t]
  \centering
  \includegraphics[width=0.80\columnwidth]{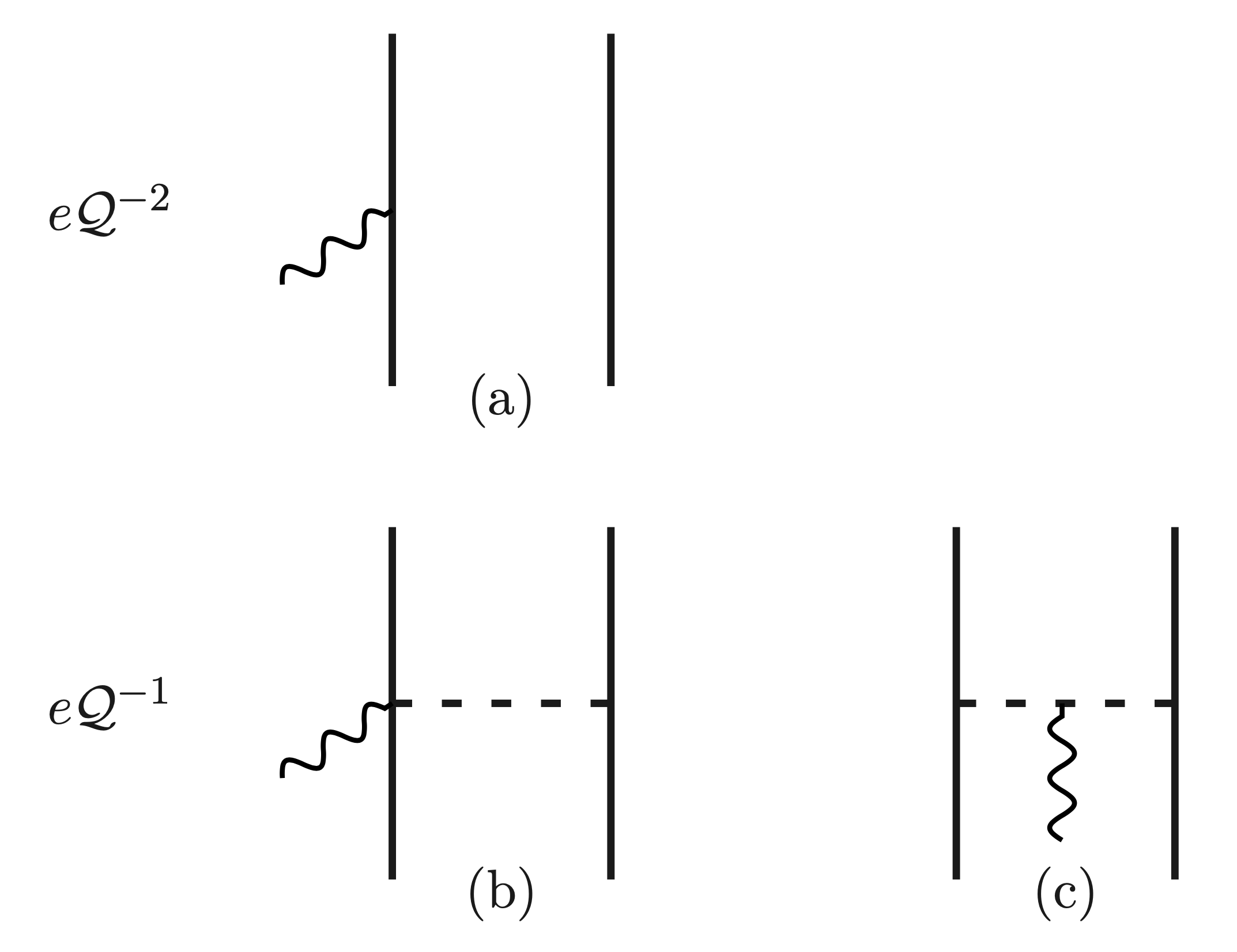}
  \caption{Diagrams for the LO (top row) and NLO (bottom row) contributions to the electromagnetic current operator, indicated by their scaling according to $e\mathcal{Q}^\nu$. Solid lines represent nucleons, while dashed and wiggly lines represent pions and photons. Diagrams (b) and (c) are the leading 2BCs given by the seagull and pion-in-flight contribution, respectively. Note that the one-body charge operator is represented by diagram (a) too, but with order $e\mathcal{Q}^{-3}$.}
  \label{fig:lo_and_nlo_current_diagrams}
\end{figure}

At NLO ($\sim e Q^{-1}$), the leading 2BC operators enter.
They are connected to the one-pion-exchange interaction and their momentum-space expressions are given by~\cite{piarulli2013}
\begin{align}
  \v{j}^{(-1)}(\v{Q}) =& -\mathrm{i} \, e \frac{g_A^2}{4F_\pi^2} G^V_\textnormal{E}(Q^2) (\boldsymbol{\tau}_1 \times \boldsymbol{\tau}_2)_z \bigg[ \boldsymbol{\sigma}_1 - \v{q}_1 \frac{\boldsymbol{\sigma}_1 \cdot \v{q}_1}{q_1^2 + m_\pi^2} \bigg] \notag \\[1mm]
  &\times \frac{\boldsymbol{\sigma}_2 \cdot \v{q}_2}{q_2^2 + m_\pi^2} + 1 \leftrightharpoons 2, \label{eq:nlo_current_operator}
\end{align}
with the axial coupling $g_A=1.27$, the pion decay constant $F_\pi=92.3$ MeV, the averaged pion mass $m_\pi=138.039$ MeV, the momentum transfers $\v{q}_i = \v{k}_i' - \v{k}_i$, the two-body center of mass momenta $\mathbf{P}^{(\prime)} = (\v{k}_1^{(\prime)} + \v{k}_2^{(\prime)})/2$, and the Pauli isospin matrices $\boldsymbol\tau_i$, operating on nucleon $i$. Here momentum conservation implies $\mathbf{P}' =\mathbf{P}+\v{Q}$.
\Cref{fig:lo_and_nlo_current_diagrams} shows the LO and NLO diagrams for the current operator.
The first and second term in the square bracket in \cref{eq:nlo_current_operator} correspond to diagram (b) and (c), respectively, which are commonly referred to as the ``seagull'' and ``pion-in-flight terms.''
Higher orders of the current operator have been derived see, e.g., Refs.~\cite{pastore2009, piarulli2013, koelling2009, koelling2011, krebs2020}.
Also, we note that the 2BC operator is not regularized in this work.
In that sense, the operator is not fully consistent with employed nuclear interactions.
Very recently, a consistent implementation was achieved by using semilocal coordinate-space regularization~\cite{Pal2023}.

For the triton (t) and helion (h), the magnetic form factor is given by~\cite{pastore2013}
\begin{align}
  F_\textnormal{M}(Q) &= -\frac{2 m_N}{Q} \langle \Psi^{\mathcal{J}, \mathcal{T}}_{\mathcal{M}'_\mathcal{J} = +\mathcal{J}, \mathcal{M}_\mathcal{T}} | \, \mathrm{j}_+(\v{Q}) \, | \Psi^{\mathcal{J}, \mathcal{T}}_{\mathcal{M}_\mathcal{J} = -\mathcal{J}, \mathcal{M}_\mathcal{T}} \rangle,
  \label{eq:form_factor_matrix_element_F_M}
\end{align}
where $Q= |\v{Q}|$, $\mathcal{J}=1/2$ and $\mathcal{M}_\mathcal{J}$ are the total three-body angular momentum and its projection, $\mathcal{T}=1/2$ and $\mathcal{M}_\mathcal{T}$ are the three-body isospin and its projection, $|\Psi \rangle$ represents the three-body state, and we have suppressed the other quantum numbers of the triton or helion. As mentioned previously, the magnetic moments of triton and helion are given by the form factors at zero momentum transfer:
\begin{equation}
\mu_{\textnormal{t}/\textnormal{h}} = F_\textnormal{M}(0).
\end{equation}

\subsection{Magnetic moment operator}

The magnetic moment operator is determined from the nuclear current operator in momentum space by~\cite{pastore2009}
\begin{equation}\label{eq:magnetic_moment_momentum}
  \boldsymbol{\mu} = -\frac{\mathrm{i}}{2} \lim_{\v{Q} \rightarrow 0} \, \boldsymbol{\nabla}_{\v{Q}} \times \v{j}(\v{Q}).
\end{equation}
The current operator can be expanded as a sum of one- and many-body operators, resulting in a similar expansion for the magnetic moment operator
\begin{equation}
  \boldsymbol{\mu} = \sum_{i}^{A} \boldsymbol{\mu}_{\textnormal{1b}, i} + \sum_{i<j}^{A} \boldsymbol{\mu}_{\textnormal{2b}, ij} + \ldots,
\end{equation}
where $\boldsymbol{\mu}_{\textnormal{1b}, i}$ is the single-nucleon contribution and $\boldsymbol{\mu}_{\textnormal{2b}, ij}$ the two-body part.
The two-body magnetic dipole operator at NLO includes contributions from diagrams (b) and (c) in \cref{fig:lo_and_nlo_current_diagrams}.

The one-body magnetic dipole operator is given by~\cite{blatt1952}
\begin{align}
\label{eq:one_body_magentic_operator}
  \boldsymbol{\mu}_{\textnormal{1b}, i} =&  \sum_{\kappa=0}^{1} (\boldsymbol{\mu}^{\kappa}_{\textnormal{spin}, i} + \boldsymbol{\mu}^{\kappa}_{\textnormal{orb}, i}), \\
  \label{eq:one_body_magentic_operator_spin0}
  \boldsymbol{\mu}^{0}_{\textnormal{spin}, i} &= G^{S}_{M}(0) \boldsymbol{\sigma}_{i}, \\
  \label{eq:one_body_magentic_operator_spin1}
  \boldsymbol{\mu}^{1}_{\textnormal{spin}, i} &= G^{V}_{M}(0) \tau_{i,z} \boldsymbol{\sigma}_{i}, \\
   \label{eq:one_body_magentic_operator_orb0}
   \boldsymbol{\mu}^{0}_{\textnormal{orb}, i} &=  \frac{\mu_{N}}{2} \boldsymbol{\ell}_{i}, \\
    \label{eq:one_body_magentic_operator_orb1}
 \boldsymbol{\mu}^{1}_{\textnormal{orb}, i} &=  \frac{\mu_{N}}{2} \tau_{i,z} \boldsymbol{\ell}_{i},
\end{align}
with the orbital angular momentum $\boldsymbol\ell_i$.

Because 2BC operators are translationally invariant with respect to the two-body center of mass $\v{R}_{ij} = (\v{r}_{i} + \v{r}_{j})/2$, the center-of-mass motion can be factored out as
\begin{equation}
  \v{j}_{{\rm 2b}, ij}(\v{Q}, \v{R}_{ij}) = e^{\mathrm{i} \v{Q} \cdot \v{R}_{ij}} \v{j}_{{\rm 2b}, ij}(\v{Q}).
\end{equation}
Accordingly, \cref{eq:magnetic_moment_momentum} splits into two parts, where one term depends only on the intrinsic coordinates and the other also on the center of mass.
The intrinsic magnetic moment operator and is then given by~\cite{pastore2009}
\begin{equation}
  \boldsymbol{\mu}_{{\rm 2b}, ij}^\textnormal{int} = -\frac{\mathrm{i}}{2} \lim_{\v{Q} \rightarrow 0} \, \boldsymbol{\nabla}_{\v{Q}} \times \v{j}_{{\rm 2b}, ij}(\v{Q}),
\end{equation}
whereas the center-of-mass dependence is contained in the so-called ``Sachs'' term~\cite{sachs1948}
\begin{equation}
  \boldsymbol{\mu}_{{\rm 2b}, ij}^\textnormal{Sachs} = \frac{1}{2} \, \v{R}_{ij} \times \v{j}_{{\rm 2b}, ij}(\v{Q}).
\end{equation}
This division into two parts can be made for 2BC operators at any order.

Summing the contributions of the seagull and the pion-in-flight terms yields for the total NLO intrinsic operator
\begin{align}\label{eq:intrinsic_total}
  \boldsymbol{\mu}^\textnormal{NLO, int}_{{\rm 2b}, ij}(\v{r}_{ij}) =& - \frac{g_A^2 m_\pi}{32 \pi F_{\pi}^2} (\boldsymbol{\tau}_i \times \boldsymbol{\tau}_j)_z \, \bigg\{ f(r_{ij}) \, \big[ (\boldsymbol{\sigma}_i \times \boldsymbol{\sigma}_j) \cdot \hat{\v{r}}_{ij} \big] \, \hat{\v{r}}_{ij} \notag \\
  &\mathopen{\hphantom{- \frac{g_A^2 m_\pi}{8 \pi F_{\pi}^2} (\boldsymbol{\tau}_1 \times \boldsymbol{\tau}_2)_z \, \bigg\{f}} - (\boldsymbol{\sigma}_i \times \boldsymbol{\sigma}_j) \bigg\} e^{-m_\pi r_{ij}},
\end{align}
with $f(r_{ij}) = 1 + 1/(m_\pi r_{ij})$ and the unit vector $\hat{\v{r}}_{ij}$ of $\v{r}_{ij} = \v{r}_{i}-\v{r}_{j}$.
The result for the NLO Sachs term is given by
\begin{equation}\label{eq:sachs_total}
  \boldsymbol{\mu}^\textnormal{NLO, Sachs}_{{\rm 2b}, ij}(\v{r}_{ij}, \v{R}_{ij}) = - \frac{1}{2} (\boldsymbol{\tau}_i \times \boldsymbol{\tau}_j)_z \, V_{1\pi}(\v{r}_{ij}) \, \v{R}_{ij} \times \v{r}_{ij},
\end{equation}
where $V_{1\pi}(\v{r}_{ij})$ is the coordinate-space one-pion-exchange potential without isospin dependence:
\begin{align}
  V_{1\pi}(\v{r}_{ij}) =& \frac{m_{\pi}^2}{12 \pi} \frac{g_A^2}{4F_{\pi}^2} \bigg\{ \bigg[ S_{ij}(\hat{\v{r}}_{ij}) \, h(r_{ij}) + \boldsymbol{\sigma}_i \cdot \boldsymbol{\sigma}_j \bigg] \frac{e^{-m_\pi r_{ij}}}{r_{ij}} \notag \\
  &\mathopen{\hphantom{\frac{m_{\pi}^2}{12 \pi} \frac{g_A^2}{F_{\pi}^2} (\boldsymbol{\tau}_1 \cdot \boldsymbol{\tau}_2) \bigg\{ \bigg[}} - \frac{4 \pi}{3} \boldsymbol{\sigma}_i \cdot \boldsymbol{\sigma}_j \, \delta(\v{r}_{ij}) \bigg\}.
\end{align}
Here, $S_{ij}(\hat{\v{r}}_{ij}) = 3 (\hat{\v{r}}_{ij} \cdot \boldsymbol{\sigma}_i)( \hat{\v{r}}_{ij} \cdot \boldsymbol{\sigma}_j) - \boldsymbol{\sigma}_i \cdot \boldsymbol{\sigma}_j$ and $h(r_{ij}) = 1 + 3/(m_\pi r_{ij}) + 3/(m_\pi r_{ij})^2$.
At NLO, the Sachs term is determined by the one-pion-exchange potential only.

\section{Operator matrix elements}\label{sec:operator_matrix_elements}

Matrix elements of operators expanded with respect to a chosen computational basis are essential components for calculating observables in few- and many-body methods.
This section focuses on expanding the one- and two-body current and magnetic dipole operators into matrix elements with respect to a specific basis.
First, we show the expansion of the current operators with respect to  one-, two-, and three-body momentum-space Jacobi bases, where the three-body result is expressed in terms of the one- and two-body matrix elements.
Next, we expand the magnetic dipole operator contributions with respect to one-, two-, and three-body relative HO bases.
As the Sachs term explicitly depends on the two-body center-of-mass coordinate, embedding it into three-body Jacobi basis needs additional steps.
We will show this embedding for the Sachs term in detail.

The single-particle partial-wave momentum-space basis states we employ are given by
\begin{align}\label{eq:single_particle_momentum_basis}
  \ket{k_i \, (\ell_i s_i) j_i m_{j_i}\, t_i m_{t_i}},
\end{align}
with the absolute value of the single-particle momentum $k_i=|\v{k}_i|$, orbital angular momentum $l_i$, spin $s_i=\tfrac{1}{2}$, total angular momentum $j_i$ and its projection $m_{j_{i}}$, and isospin $t_i = \tfrac{1}{2}$ along with its projection $m_{t_i}$ for nucleon $i$.
Relative two-body quantum numbers are denoted by capital letters and relative two-body momentum-space basis states are defined by
\begin{align}\label{eq:two_body_momentum_basis}
  \ket{p \, \alpha_\text{2b}} \equiv \ket{p \, (L S)J M_J \, T M_T},
\end{align}
with the relative momentum $\mathbf{p} = \frac{1}{2} (\mathbf{k}_1 - \mathbf{k}_2)$ of two nucleons, $p=|\v{p}|$, the relative orbital angular momentum $L$, two-body spin $S$, total angular momentum $J$ and its projection $M_{J}$, and total isospin $T$ and its projection $M_{T}$.
The collective index $\alpha_\text{2b}$ defines the set of two-body quantum numbers $\alpha_\text{2b} = \{ L, S, J, T \}$.
Three-body basis states are constructed by coupling a third nucleon to the two-body system and defining it relative to the center of mass of the nucleon pair with momentum $\mathbf{p}$:
\begin{align}\label{eq:three_body_momentum_basis}
  \ket{p \, q \, \alpha} \equiv \ket{p \, q \, [(L S)J \, (\ell s)j]\mathcal{J}\mathcal{M}_\mathcal{J} \, (T t) \mathcal{T} \mathcal{M}_\mathcal{T}}.
\end{align}
Here $\v{q} = \frac{2}{3} \left( \mathbf{k}_3 - \frac{1}{2} (\mathbf{k}_1 + \mathbf{k}_2) \right)$ is the second Jacobi momentum, whereas $\ell$, $s$, $j$, and $t$ are the corresponding spin, isospin and angular momentum quantum numbers~\cite{Hebeler:2020ocj}.
The total three-body angular momentum and isospin are denoted by $\mathcal{J}$ and $\mathcal{T}$, respectively.
Here again, the collective index $\alpha = \{ L, S, J, T, l, s, j, \mathcal{J}, \mathcal{T} \}$ contains the partial-wave quantum numbers that define the state.

Partial-wave HO states are constructed in a similar manner, with the only difference being that the momentum is exchanged by the principle HO quantum number.
Accordingly, the single-particle HO basis states are given by
\begin{align}\label{eq:single_particle_ho_basis}
  \ket{n_i (\ell_i s_i) j_i m_{j_i} \, t_i m_{t_i}},
\end{align}
while the two-body basis states become
\begin{align}\label{eq:two_body_ho_basis}
  \ket{N \, \alpha_\text{2b}} \equiv \ket{N (L S) J M_J \, T M_T},
\end{align}
and the three-body HO basis states are specified by
\begin{align}\label{eq:three_body_ho_basis}
  \ket{N \, n \, \alpha} \equiv \ket{N \, n \, [(L S)J \, (\ell s)j]\mathcal{J}\mathcal{M}_{\mathcal{J}} \, (T t) \mathcal{T} \mathcal{M}_{\mathcal{T}}}.
\end{align}
Note that the Jacobi coordinates used in our NCSM calculations are not exactly the same as those used in the Faddeev calculations. All definitions are provided in Appendix~\ref{sec:Jacobi_coodinate}.

\subsection{Partial-wave expanded current operator}\label{subsec:current_operator_expansion}
Generally, the matrix elements of one-body and two-body current operators, as defined in Eqs.~(\ref{eq:lo_current_operator}) and (\ref{eq:nlo_current_operator}), can be expressed in the following form within the three-body partial-wave basis defined in Eq.~(\ref{eq:three_body_momentum_basis})~\cite{Golak_2005,Hebeler:2020ocj}:
\begin{widetext}
\begin{align}
\left\langle p^{\prime} q^{\prime} \alpha^{\prime}\left| \mathbf{j} (\mathbf{Q})\right| p q \alpha\right\rangle= & \sum_{M_J M_J^{\prime} m_j m_j^{\prime}}
\mathcal{C}_{J' M'_J j m'_j}^{\mathcal{J}^{\prime} \mathcal{M}_{\mathcal{J}}^{\prime}}
\mathcal{C}_{J M_J j m_j}^{\mathcal{J} \mathcal{M}_{\mathcal{J}}} \sum_{M_T M'_T m_t m'_t}
\mathcal{C}_{T' M'_T t' m'_t}^{\mathcal{T}' \mathcal{M}'_{\mathcal{T}}}
\mathcal{C}_{T M_T t m_t}^{\mathcal{T} \mathcal{M}_{\mathcal{T}}} \mathcal{P}_{\left(L^{\prime} S^{\prime}\right) J^{\prime} T^{\prime}(L S) J T}^{M_J^{\prime} M_T^{\prime} M_J M_T} \left(\mathbf{Q}, p, p^{\prime}\right) \mathcal{Q}_{\left(\ell^{\prime} s^{\prime} \right) j^{\prime} t^{\prime}(\ell s) j t}^{m_j^{\prime} m_t^{\prime} m_j m_t} \left(\mathbf{Q}, q, q^{\prime}\right),
\end{align}
with
\begin{align}
\mathcal{P}_{\left(L^{\prime} S^{\prime}\right) J^{\prime} T^{\prime}(L S) J T}^{M_J^{\prime} M_T^{\prime} M_J M_T} \left(\mathbf{Q}, p, p^{\prime}\right)= & \frac{1}{(2 \pi)^3} \int \mathrm{d} \mathbf{p}_1 \mathrm{d} \mathbf{p}_1^{\prime} \frac{\delta\left(p_1^{\prime}-p^{\prime}\right)}{p_1^{\prime} p^{\prime}} \mathcal{Y}_{L^{\prime} S^{\prime}}^{* J^{\prime} M_J^{\prime}}\left(\hat{\mathbf{p}}_1^{\prime}\right) \left\langle\mathbf{p}_1^{\prime} T^{\prime} M_T^{\prime}\left| \mathbf{j} (\mathbf{Q})\right| \mathbf{p}_1 T M_T \right\rangle \frac{\delta\left(p-p_1\right)}{p p_1} \mathcal{Y}_{L S}^{J M_J}\left(\hat{\mathbf{p}}_1\right), \\
\mathcal{Q}_{\left(l^{\prime} s^{\prime}\right) j^{\prime} t^{\prime}(\ell s) j t}^{m_j^{\prime} m_t^{\prime} m_j m_t}\left(\mathbf{Q}, q, q^{\prime}\right)= & \frac{1}{(2 \pi)^3} \int \mathrm{d} \mathbf{q}_1 \mathrm{d} \mathbf{q}_1^{\prime} \frac{\delta\left(q_1^{\prime}-q^{\prime}\right)}{q_1^{\prime} q^{\prime}} \mathcal{Y}_{\ell^{\prime} s^{\prime}}^{* j_j^{\prime} m'_j}\left(\hat{\mathbf{q}}_1^{\prime}\right) \left\langle\mathbf{q}_1^{\prime} t^{\prime} m_t^{\prime}\left| \mathbf{j} (\mathbf{Q})\right| \mathbf{q}_1 t m_t\right\rangle \frac{\delta\left(q-q_1\right)}{q q_1} \mathcal{Y}_{\ell s}^{j m_j}\left(\hat{\mathbf{q}}_1\right) ,
\end{align}
\end{widetext}
and the spinor spherical harmonics
\begin{align}
\mathcal{Y}_{L S}^{J M_J} \left(\hat{\mathbf{a}}\right) = \sum_{M_L, M_S} \mathcal{C}_{L M_L S M_S}^{J M_J} Y_{L M_L} \left(\hat{\mathbf{a}} \right) \left| S M_S \right> .
\end{align}
This factorized representation is very useful in practice as one-body operator can be represented in a natural way in terms of the quantity $\mathcal{Q} \left(\mathbf{Q}, q, q^{\prime}\right)$, while for two-body operators the dynamics of the interaction with the external probe can be parametrized naturally via $\mathcal{P} \left(\mathbf{Q}, p, p^{\prime}\right)$.

\subsubsection{One-body operators}

For the representation of one-body currents it is convenient to choose the coordinates such that the external probe interacts with the ``last'' particle, which for the 3N system is the nucleon with Jacobi momentum $\mathbf{q}$ resp.~$\mathbf{q}'$ describing its motion relative to the subsystem of the other two nucleons. Specifically, we have:
\begin{align}
\mathbf{k}'_1 = \mathbf{k}_1, \quad \mathbf{k}'_2 = \mathbf{k}_2, \quad \mathbf{k}'_3 = \mathbf{k}_3 + \mathbf{Q} ,
\end{align}
\begin{widetext}
which implies for the Jacobi momenta:
\begin{equation}
\mathbf{p}' = \mathbf{p}, \quad
\mathbf{q}' = \mathbf{q} + \frac{2}{3} \mathbf{Q} \,.
\label{eq:OBC_momenta}
\end{equation}
Consequently, the matrix elements of $\mathcal{P} (\mathbf{Q}, p, p')$ are trivial since both particles with the relative momentum $\mathbf{p} = \mathbf{p}'$ are just spectator particles:
\begin{align}
    &\mathcal{P}_{\left(L^{\prime} S^{\prime}\right) J^{\prime} T^{\prime}(L S) J T}^{M_J^{\prime} M_T^{\prime} M_J M_T} \left(\mathbf{Q}, p, p^{\prime}\right) = \delta_{M_T M_T^{\prime}} \delta_{M_J M'_J}\delta_{J J'}  \delta_{L L^{\prime}} \delta_{S S^{\prime}} \delta_{T T^{\prime}} \frac{\delta\left(p-p^{\prime}\right)}{p^{\prime} p} \, .
\end{align}
The quantity $\mathcal{Q} (\mathbf{Q}, q, q')$ on the other hand contains the dynamics of the probe described by the one-body current and is given by:
\begin{align}
\mathcal{Q}_{\left(l^{\prime} s^{\prime}\right) j^{\prime} t^{\prime}(\ell s) j t}^{m_j^{\prime} m_t^{\prime} m_j m_t}\left(\mathbf{Q}, q, q^{\prime}\right) = \int \mathrm{d} \hat{\mathbf{q}} \frac{\delta \left(q' - |\mathbf{q} + \frac{2}{3} \mathbf{Q}| \right)}{q'^2} \mathcal{Y}_{\ell^{\prime} s^{\prime}}^{* j_j^{\prime} m'_j}\left(\widehat{\mathbf{q} + \frac{2}{3} \mathbf{Q}} \right) \left\langle (\mathbf{q} + 2/3 \mathbf{Q}) t^{\prime} m_t^{\prime}\left| \mathbf{j} (\mathbf{Q})\right| \mathbf{q} t m_t\right\rangle \mathcal{Y}_{\ell s}^{j m_j}\left(\hat{\mathbf{q}}\right) \, .
\end{align}
The evaluation is straightforward and can be performed numerically or partially analytically, depending on the specific form of the current operator.

\subsubsection{Two-body operators}
In the case of 2BC operators it is most convenient for the practical evaluation of the matrix elements to choose the two-body subsystem characterized by the Jacobi momenta $\mathbf{p}$ resp.~$\mathbf{p}'$ to interact with the external probe, i.e., $\mathbf{k}'_3 = \mathbf{k}_3$. We first express the single-particle momenta in terms of the Jacobi and center-of-mass momenta \cite{Hebeler:2020ocj}:
\begin{align}
    \mathbf{k}_1 &= \mathbf{p} - \frac{\mathbf{q}}{2} + \frac{1}{3} \mathbf{P}_{\mathrm{3N}} \\
    \mathbf{k}_2 &= - \mathbf{p} - \frac{\mathbf{q}}{2} + \frac{1}{3} \mathbf{P}_{\mathrm{3N}} \\
  \mathbf{k}_3 &= \mathbf{q} + \frac{1}{3} \mathbf{P}_{\mathrm{3N}}.
\end{align}
As a result of the interaction process, the center-of-mass momentum changes, $\mathbf{P}'_{\mathrm{3N}} = \mathbf{P}_{\mathrm{3N}} + \mathbf{Q}$, and hence:
\begin{align}
\mathbf{q}' = \mathbf{q} + \frac{1}{3} \mathbf{Q},
\end{align}
while the Jacobi momentum $\mathbf{p} = (\mathbf{k}_2 - \mathbf{k}_1)/2$ and $\mathbf{p}' = (\mathbf{k}'_2 - \mathbf{k}'_1)/2$ are again unaffected by the interaction with the external probe. Overall, we obtain:
\begin{align}
\mathcal{P}_{\left(L^{\prime} S^{\prime}\right) J^{\prime} T^{\prime}(L S) J T}^{M_J^{\prime} M_T^{\prime} M_J M_T} \left(\mathbf{Q}, p, p^{\prime}\right) &= \frac{1}{(2 \pi)^3} \int \mathrm{d} \hat{\mathbf{p}}^{\prime} d \hat{\mathbf{p}} \, \mathcal{Y}_{L^{\prime} S^{\prime}}^{* J^{\prime} M_J^{\prime}}\left(\hat{\mathbf{p}}^{\prime}\right) \left\langle\mathbf{p}^{\prime} T^{\prime} M_T^{\prime}\left| \mathbf{j} (\mathbf{Q})\right| \mathbf{p} T M_T\right\rangle \mathcal{Y}_{L S}^{J M_J}\left(\hat{\mathbf{p}} \right) \\
\mathcal{Q}_{\left(l^{\prime} s^{\prime}\right) j^{\prime} t^{\prime}(\ell s) j t}^{m_j^{\prime} m_t^{\prime} m_t m_t}\left(\mathbf{Q}, q, q^{\prime}\right) &= \delta_{m_t m'_t} \delta_{s s'} \delta_{m_s m'_s} \int \mathrm{d} \hat{\mathbf{q}} \frac{\delta \left(q' - |\mathbf{q} + \frac{1}{3} \mathbf{Q}| \right)}{q'^2} \mathcal{Y}_{\ell^{\prime} s^{\prime}}^{* j_j^{\prime} m_j}\left(\widehat{\mathbf{q} + \frac{1}{3} \mathbf{Q}} \right) \mathcal{Y}_{\ell s}^{j m_j}
\left(\hat{\mathbf{q}}\right) \, .
\end{align}
\end{widetext}
The quantity $\mathcal{Q}$ is a current-independent function that in this case depends only on the kinematics specified by the momentum $\mathbf{Q}$, while the two-body quantity $\mathcal{P}$ contains all the information about the current operator.

\subsection{Harmonic-oscillator expanded magnetic dipole operator}\label{subsec:dipole_operator_expansion}
In this section, we show matrix elements expressed in the HO basis, which enter the three-body Jacobi NCSM calculations.
Without loss of generality we choose the external momentum $\v{Q}$ to be along the $z$ direction.
In the Jacobi NCSM, a wave function $|\Psi_{\rm 3b}\rangle$ is obtained through diagonalization of the Hamiltonian and expressed as a superposition of antisymmetrized HO basis states:
\begin{equation}
|\Psi_{\rm 3b} \rangle = \sum_{i} c_{i} |i \rangle.
\end{equation}
Note that the antisymmetrized HO basis $|i\rangle$ is not the same as the state defined in Eq.~\eqref{eq:three_body_ho_basis} and computed as
\begin{equation}
| i \rangle = \sum_{Nn\alpha} |Nn\alpha \rangle \langle Nn\alpha | i \rangle,
\end{equation}
with the coefficient of fractional parentage $\langle Nn\alpha| i \rangle$~\cite{Navratil1998,Barrett2013}.
Through the antisymmetrization, it is clear that expectation values do not depend on the choice of the three-body Jacobi coordinate, i.e., one can choose the spectator particle.
For example, the basis definitions~\eqref{eq:three_body_momentum_basis} and  \eqref{eq:three_body_ho_basis} take spectator particle as the third one.
Exploiting this, one can find
\begin{multline}
\langle \Psi_{\rm 3b}' | \sum_{i}\mu_{{\rm 1b}, i} |\Psi_{\rm 3b} \rangle =
3 \sum_{kl} \sum_{N'n'\alpha'} \sum_{Nn\alpha} c'^{*}_{k} c_{l}
\\  \times
\langle k | N'n'\alpha' \rangle
\langle Nn\alpha | l \rangle
\langle N'n'\alpha' | \mu_{{\rm 1b}, 3} | Nn\alpha \rangle,
\end{multline}
for one-body operators.
A similar expression can be found for two-body operators:
\begin{multline}
\langle \Psi_{\rm 3b}' | \sum_{i<j}\mu_{{\rm 2b}, ij} |\Psi_{\rm 3b} \rangle =
3 \sum_{kl} \sum_{N'n'\alpha'} \sum_{Nn\alpha} c'^{*}_{k} c_{l}
\\ \times
\langle k | N'n'\alpha' \rangle
\langle Nn\alpha | l \rangle
\langle N'n'\alpha' | \mu_{{\rm 2b}, 12} | Nn\alpha \rangle.
\end{multline}
The main tasks are to find expressions for
\begin{equation}
\label{eq:mu1_with_3body_ho}
\langle N'n'\alpha' ||| \mu^{\kappa}_{{\rm 1b}, 3} ||| Nn\alpha \rangle,
\end{equation}
and
\begin{equation}
\label{eq:mu2_with_3body_ho}
\langle N'n'\alpha' ||| \mu^{\kappa}_{{\rm 2b}, 12} ||| Nn\alpha \rangle.
\end{equation}
Here, we introduced doubly reduced matrix element with respect to spin and isospin, where $\kappa$ is the isospin rank of the magnetic moment operator.
This does not lose any information, and one can always restore normal matrix elements by means of the Wigner-Eckart theorem.

\subsubsection{Harmonic-oscillator basis}

To compute matrix elements within the HO basis, we first define the momentum-space representation of radial oscillator wave functions for a single-particle $\tilde{R}_{n\ell}(k) = \braket{k \, \ell}{n \, \ell}$, defined by the overlap between momentum and HO eigenstates, given by the slightly modified definition from Ref.~\cite{caprio2012}
\begin{equation}
  \tilde{R}_{n\ell}(k) = \sqrt{\frac{2n! b^3}{\Gamma(n + \ell + 3/2)}} \, \big(kb\big)^\ell \, e^{-\frac{1}{2} k^2 b^2} \, L_n^{\ell + 1/2}(k^2 b^2),
\end{equation}
with $b \equiv 1/\sqrt{m_N\omega}$ the oscillator length in terms of the oscillator frequency $\omega$ and the nucleon mass $m_N$, and $L_n^\ell(x)$ are generalized Laguerre polynomials.
Similarly, coordinate-space radial wave functions $R_{n\ell}(r) = \braket{r \, \ell}{n \, \ell}$ are given by
\begin{align}
  R_{n\ell}(r) =& (-1)^n \sqrt{\frac{2n!}{\Gamma(n + \ell + 3/2)b^3}} \, \Bigg(\frac{r}{b}\Bigg)^\ell \, e^{-\frac{1}{2}(\frac{r}{b})^2} L_n^{\ell + 1/2}\Bigg(\frac{r^2}{b^2}\Bigg),
\end{align}
which are connected to the momentum-space functions through a Fourier-Bessel transform
\begin{align}
  R_{n\ell}(r) = \int\!\mathrm{d}k \, k^2 \braket{r \, \ell}{k \, \ell} \tilde{R}_{n\ell}(k),
\end{align}
where the overlap is described by spherical Bessel functions $\braket{r \, \ell}{k \, \ell}=\sqrt{2/\pi}j_\ell(kr)$.

\subsubsection{One-body operator}

Matrix elements of the one-body magnetic dipole operator defined in Eqs.~\eqref{eq:one_body_magentic_operator_spin0}-\eqref{eq:one_body_magentic_operator_orb1} are given by
\begin{align}
\langle n'\ell'j' ||| \mu^{0}_{\rm spin} ||| n\ell j \rangle &=  2\sqrt{3} G^{S}_{M}(0)\hat{j}'\hat{j} (-1)^{\ell'+j'+3/2}
 \notag \\  & \quad \times \sixj{1/2}{1/2}{1}{j'}{j}{\ell} \delta_{nn'}\delta_{\ell\ell'}, \\
 \langle n'\ell'j' ||| \mu^{1}_{\rm spin} ||| n\ell j \rangle &=  6 G^{V}_{M}(0)\hat{j}'\hat{j} (-1)^{\ell'+j'+3/2}
 \notag \\  & \quad \times \sixj{1/2}{1/2}{1}{j'}{j}{\ell} \delta_{nn'}\delta_{\ell\ell'}, \\
 \langle n'\ell'j' ||| \mu^{0}_{\rm orb} ||| n\ell j \rangle &=  \frac{\mu_{N}}{\sqrt{2}} \hat{j}'\hat{j} \hat{\ell}
 \sqrt{\ell(\ell+1)} (-1)^{\ell+j+3/2} \notag \\  & \quad \times \sixj{\ell}{\ell}{1}{j'}{j}{1/2} \delta_{nn'}\delta_{\ell\ell'}, \\
 \langle n'\ell'j' ||| \mu^{1}_{\rm orb} ||| n\ell j \rangle &=  \sqrt{\frac{3}{2}} \mu_{N} \hat{j}'\hat{j} \hat{\ell}
 \sqrt{\ell(\ell+1)} (-1)^{\ell+j+3/2} \notag \\  & \quad \times \sixj{\ell}{\ell}{1}{j'}{j}{1/2} \delta_{nn'}\delta_{\ell\ell'}.
\end{align}

For the spin term $\mu^{\kappa}_{\rm spin}$, the required reduced matrix element for Eq.~\eqref{eq:mu1_with_3body_ho} is
\begin{equation}
\begin{aligned}
\langle N'n'\alpha ||| &\mu^{\kappa}_{{\rm spin},3} ||| Nn\alpha \rangle = (-1)^{\mathcal{J}'+J+j+1} \hat{\mathcal{J}}' \hat{\mathcal{J}} \sixj{j'}{\mathcal{J}'}{J}{\mathcal{J}}{j}{1}
\\ &\times
(-1)^{\mathcal{T}'+T+3/2} \hat{\mathcal{T}}' \hat{\mathcal{T}} \sixj{1/2}{\mathcal{T}'}{T}{\mathcal{T}}{1/2}{\kappa}
\\ & \times
\langle n'\ell'j' ||| \mu^{\kappa}_{\rm spin} ||| n\ell j \rangle
\delta_{N'N} \delta_{L'L} \delta_{S'S} \delta_{J'J}\delta_{T'T}.
\end{aligned}
\end{equation}
A similar expression can be found for the orbital contribution $\mu^{\kappa}_{\rm orb}$:
\begin{equation}
\begin{aligned}
\langle N'n'\alpha ||| & \mu^{\kappa}_{{\rm orb},3} ||| Nn\alpha \rangle = \frac{2}{3} (-1)^{\mathcal{J}'+J+j+1} \hat{\mathcal{J}}' \hat{\mathcal{J}} \sixj{j'}{\mathcal{J}'}{J}{\mathcal{J}}{j}{1}
\\ &\times
(-1)^{\mathcal{T}'+T+3/2} \hat{\mathcal{T}}' \hat{\mathcal{T}} \sixj{1/2}{\mathcal{T}'}{T}{\mathcal{T}}{1/2}{\kappa}
\\ & \times
\langle n'\ell'j' ||| \mu^{\kappa}_{\rm orb} ||| n\ell j \rangle
\delta_{N'N} \delta_{L'L} \delta_{S'S} \delta_{J'J}\delta_{T'T}.
\end{aligned}
\end{equation}
Notice that there is a factor $2/3$, coming from the transformation from single-particle to three-body Jacobi coordinates.
Also, we have used that we can choose the orbital angular momentum of the three-body center-of-mass coordinate as zero since the intrinsic and center-of-mass motions do not couple.

\subsubsection{Intrinsic magnetic dipole operator}
After angular momentum recoupling, the reduced matrix element of the intrinsic magnetic dipole operator \eqref{eq:intrinsic_total} with respect to a relative two-body HO basis \eqref{eq:two_body_ho_basis} can be computed through
\begin{equation}
\begin{aligned}
\langle & N'\alpha'_{\rm 2b} ||| \mu^{\rm NLO, int}_{\rm 2b} ||| N\alpha_{\rm 2b} \rangle  =
 -\frac{g^{2}_{A}m_{\pi}}{32\pi F^{2}_{\pi}} \sqrt{\frac{8\pi}{3}}
 \\ & \times
( i \langle T' || (\boldsymbol{\tau}_{1} \times \boldsymbol{\tau}_{2}) || T \rangle )
 \\ & \times
 \int dr r^{2} R_{N'L'}(r) R_{NL}(r) e^{-\sqrt{2} m_{\pi}r}
 \\ & \times
 \sum_{w=0,2} A^{L'S'J'}_{LSJ}(w,1)
 \left[
 f(\sqrt{2} r)  \clebsch{1}{0}{1}{0}{w}{0}
 + \sqrt{3} \delta_{w0}
 \right],
 \label{eq:intrinsic_two_body_expansion}
\end{aligned}
\end{equation}
with
\begin{equation}
\begin{aligned}
A^{L'S'J'}_{LSJ}(w,x) &\equiv \langle (L'S')J' || [Y_{w}(\hat{\v{r}})[\sigma_{1}\sigma_{2}]_{x}]_{1} || (LS)J \rangle \\
& = 6 \sqrt{\frac{3}{4\pi}} \hat{J'}\hat{J}\hat{L}\hat{S'}\hat{S}\hat{w}\hat{x}
\\ & \quad \times \ninej{L'}{L}{w}{S'}{S}{x}{J'}{J}{1} \ninej{1/2}{1/2}{1}{1/2}{1/2}{1}{S'}{S}{x}
\clebsch{L}{0}{w}{0}{L'}{0}.
\end{aligned}
\end{equation}
Also, the matrix element of the intrinsic two-body magnetic moment operator for Eq.~\eqref{eq:mu2_with_3body_ho} can be written as
\begin{equation}
\begin{aligned}
\langle N'n'\alpha' ||| & \mu^{\rm NLO, int}_{{\rm 2b},12} ||| Nn\alpha \rangle =
  (-1)^{J'+j'+\mathcal{J}+1}(-1)^{T'+\mathcal{T}+3/2}
  \\ &  \times
\hat{\mathcal{J}}'\hat{\mathcal{J}}
\sixj{\mathcal{J}'}{J'}{j}{J}{\mathcal{J}}{1}
\\ &  \times
\hat{\mathcal{T}}'\hat{\mathcal{T}}
\sixj{\mathcal{T}'}{T'}{1/2}{T}{\mathcal{T}}{1}
\\ & \times
\langle N'\alpha'_{2b} ||| \mu^{\rm NLO,int}_{\rm 2b} ||| N\alpha_{2b} \rangle
\delta_{n'n}\delta_{\ell' \ell}\delta_{j'j}.
\end{aligned}
\end{equation}

\subsubsection{Sachs operator}
In the previous section, we described how to embed a two-body operator that only depends on the relative coordinate between two nucleons with respect to a relative two-body basis and a three-body Jacobi basis.
Here, we consider a two-body operator depending on the two-nucleon center-of-mass, in addition to the relative coordinate.
The Sachs contribution to the NLO magnetic dipole operator is of this type.

\begin{figure}[h!]
\centering
  \includegraphics[width=\columnwidth]{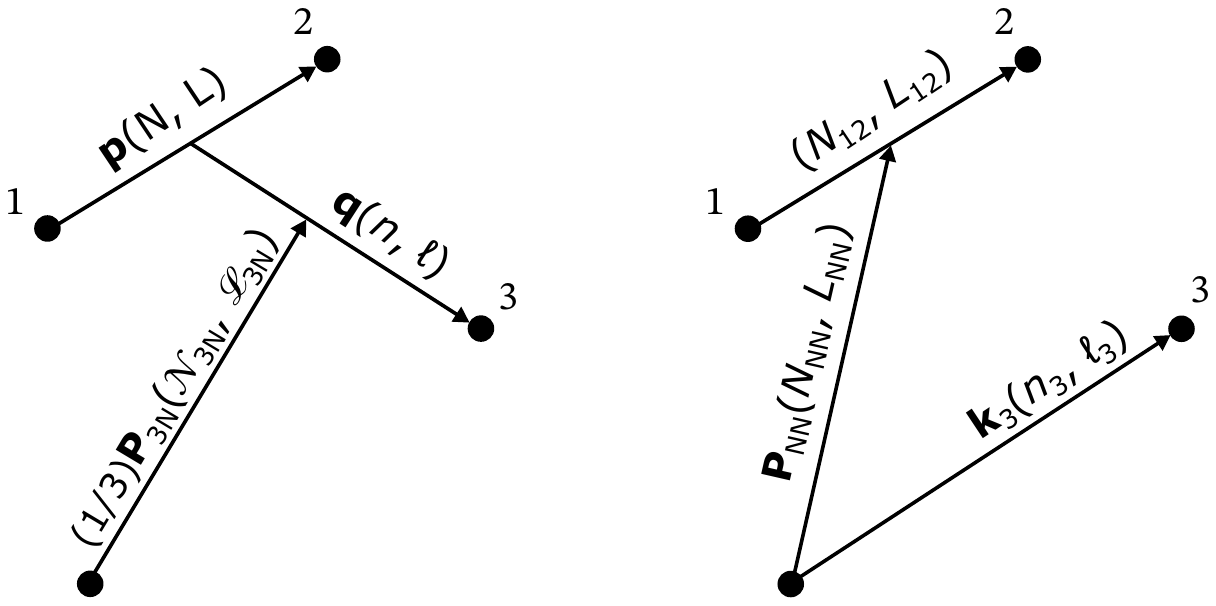}
  \caption{Schematic of two different coordinate systems representing a three-nucleon system. The left part represents the Jacobi coordinate system characterized by ($\v{P}_\text{3N}$, $\v{p}$, $\v{q}$), while the right part shows the coordinate system described by ($\v{P}_\text{NN}$, $\v{p}$, $\v{k}_3$).}
  \label{fig:coordinate_systems}
\end{figure}

We start by constructing basis states that explicitly include the two-body center-of-mass motion.
Such a basis is denoted by
\begin{multline}
   \ket{N_\text{NN} N_{12}  \, \alpha_{12}} \\
  \equiv  \ket{N_\text{NN}N_{12}  [L_\text{NN} (L_{12} S_{12})J_{12} ] J_\text{rc} M_{J_\text{rc}} T_{12} M_{T_{12}}}, \label{eq:two_body_center_of_mass_basis}
\end{multline}
where the subscript ${12}$ expresses relative quantities between nucleon $1$ and $2$, while $\text{NN}$ indicates quantities related to the two-body center of mass.
Coupling the relative and center-of-mass angular momenta generates the total angular momentum of the two-body system $J_\text{rc}$ with projection $M_{J_\text{rc}}$.
A schematic representation of this basis is displayed in the right half of \cref{fig:coordinate_systems}, which shows the momenta associated to a two-body system with respect to the origin by black dots labeled $1$ and~$2$.\\

The matrix element of the Sachs operator with respect to the states defined in \cref{eq:two_body_center_of_mass_basis} is given by
\begin{equation}
\begin{aligned}
\label{eq:sachs_two_body_me}
\langle & N'_{\rm NN}N'_{12}\alpha'_{12} ||| \mu^{\rm NLO, Sachs}_{\rm 2b} ||| N_{\rm NN} N_{12}\alpha_{12} \rangle = \frac{g^{2}_{A}m^{2}_{\pi}}{24F^{2}_{\pi}}
\\ & \times
i \langle T' || (\boldsymbol{\tau}_{1} \times \boldsymbol{\tau}_{2}) || T \rangle
\\ & \times
\int dr r^{2} R_{N'_{12}L'_{12}}(r) R_{N_{12}L_{12}}(r) e^{-\sqrt{2}m_{\pi}r}
\\ & \times
\sum_{x=0,2} \sum_{w}
\left[\left(\clebsch{1}{0}{1}{0}{x}{0} \clebsch{1}{0}{x}{0}{w}{0} + \sqrt{\frac{1}{3}}\delta_{w1}\delta_{x0} \right)h(\sqrt{2} r) - \sqrt{\frac{1}{3}}\delta_{w1}\delta_{x0}  \right]
\\ & \times
\int dR R^{3} R_{N_{\rm NN}'L_{\rm NN}'}(R) R_{N_{\rm NN}L_{\rm NN}}(R)
B^{L_{\rm NN}' L_{12}'S_{12}'J_{12}'J_{\rm rc}'}_{L_{\rm NN} L_{12}S_{12}J_{12}J_{\rm rc}}(w,x),
\end{aligned}
\end{equation}
with
\begin{widetext}
\begin{equation}
\begin{aligned}
B^{L_{\rm NN}' L_{12}'S_{12}'J_{12}'J_{\rm rc}'}_{L_{\rm NN} L_{12}S_{12}J_{12}J_{\rm rc}}(w,x)  & \equiv
\langle [L_{\rm NN}' (L_{12}'S_{12}')J_{12}']J_{\rm rc}' || [Y_{1}(\hat{\v{R}})[Y_{w}(\hat{\v{r}}) [\sigma_{1}\sigma_{2}]_{x}]_{1}]_{1}|| [L_{\rm NN} (L_{12}S_{12})J_{12}]J_{\rm rc}'\rangle \\
&= 3 \sqrt{\frac{1}{4\pi}} \hat{J}_{\rm rc}' \hat{J}_{\rm rc} \hat{L}_{\rm NN} \ninej{L_{\rm NN}'}{L_{\rm NN}}{1}{J_{12}'}{J_{12}}{1}{J_{\rm rc}'}{J_{\rm rc}}{1} \clebsch{L_{\rm NN}}{0}{1}{0}{L_{\rm NN}'}{0}
A^{L_{12}'S_{12}'J_{12}'}_{L_{12}S_{12}J_{12}}(w,x).
\end{aligned}
\end{equation}
\end{widetext}
In addition to the integration over $r$, the matrix elements of the Sachs operator are integrated over the center-or-mass coordinate $R$.
Same as the intrinsic contribution, the isospin rank of the operator $\kappa$ is $1$.

Finding an expression for the matrix element of the Sachs operator in terms of three-body Jacobi basis states requires more work compared to the matrix elements we evaluated in previous sections.
First, the two-body basis we defined in \cref{eq:two_body_center_of_mass_basis} has to be extended to include a third nucleon.
This is achieved by coupling the total two-body angular momentum and isospin to the total angular momentum and isospin of the third nucleon to obtain three-body quantities.
Second, this basis allows to evaluate the Sachs operator in terms of three-body states which include the two-body center-of-mass motion, so that the result in \cref{eq:sachs_two_body_me} can be used to express the matrix element.
This expression, however, is unsuitable to calculate expectation values of the operator because the NCSM wave functions are expressed with three-body Jacobi states.
Therefore, in a third step, we determine the overlap between the two different three-body bases.

To consider three particles in a basis which includes the two-body center-of-mass motion, we add a third nucleon represented by $\ket{n_3 (\ell_3 \tfrac{1}{2}) j_3}$, which is defined with respect to the origin, and couple its angular momentum $j_3$ with the total two-body angular momentum $J_\text{rc}$ to the total three-body angular momentum $\mathcal{J}_\text{tot}$:
\begin{multline}
  \ket{N_\text{NN} N_{12}  n_3 \, \alpha_{12} \, \alpha_3 \, \mathcal{J}_\text{tot} \mathcal{M}_{\mathcal{J}_\text{tot}} \mathcal{T} \mathcal{M}_\mathcal{T}} \\
  \begin{split}
  \equiv &  \ket{N_\text{NN} N_{12}  n_3 \{[L_\text{NN}(L_{12} S_{12})J_{12} ] J_\text{rc} (\ell_3 \tfrac{1}{2}) j_{3}\} \mathcal{J}_\text{tot} \mathcal{M}_{\mathcal{J}_\text{tot}}} \\
  & \times\ket{(T_{12} \tfrac{1}{2}) \mathcal{T} \mathcal{M}_\mathcal{T}}.
  \end{split}\label{eq:three_body_com_ho_basis}
\end{multline}
The right part of \cref{fig:coordinate_systems} shows the coordinate system that corresponds to this basis state.

The matrix elements of the Sachs operator using the states defined in Eq.~\eqref{eq:three_body_com_ho_basis} can be represented in terms of reduced two-body matrix elements from \cref{eq:sachs_two_body_me}, angular momentum and isospin coupling factors, and a third particle which is diagonal in all its quantum numbers:
\begin{equation}
\begin{aligned}
\langle & N'_{\rm NN}N'_{12}  n'_{3}\alpha'_{12}\alpha'_{3} \mathcal{J}'_{\rm tot}\mathcal{T}' ||| \mu^{\rm NLO, Sachs}_{{\rm 2b},12} ||| N_{\rm NN} N_{12}  n_{3}\alpha_{12}\alpha_{3} \mathcal{J}_{\rm tot}\mathcal{T} \rangle
\\ &=
(-1)^{J'_{\rm rc}+j'_{3}+\mathcal{J}_{\rm tot}+1}
\hat{\mathcal{J}}'_{\rm tot}\hat{\mathcal{J}}_{\rm tot}
\sixj{\mathcal{J}'_{\rm tot}}{J'_{\rm rc}}{j_{3}}{J_{\rm rc}}{\mathcal{J}_{\rm tot}}{1}
\\ & \times
(-1)^{T'+\mathcal{T}+3/2}
\hat{\mathcal{T}}'\hat{\mathcal{T}}
\sixj{\mathcal{T}'}{T'}{1/2}{T}{\mathcal{T}}{1}
\\ & \times
\langle N'_{\rm NN} N'_{12} \alpha'_{12}  ||| \mu^{\rm NLO, Sachs}_{\rm 2b} ||| N_{\rm NN}N_{12}  \alpha_{12} \rangle
\delta_{n'_{3}n_{3}}\delta_{\ell'_{3}\ell_{3}}\delta_{j'_{3}j_{3}}
\label{eq:sachs_trhee_body_me}
\end{aligned}
\end{equation}
This result shows that the majority of the work to calculate three-body matrix elements consists in determining the two-body matrix elements.

To calculate the overlap between the three-body basis states defined in \cref{eq:three_body_com_ho_basis} and the three-body Jacobi states from \cref{eq:three_body_ho_basis}, the three-body center-of-mass motion has to be included.
This is done by coupling the three-body angular momentum $\mathcal{J}$ with the center-of-mass orbital angular momentum $\mathcal{L}_\text{3N}$ to the total angular momentum $\mathcal{J}_\text{tot}$, so that the basis from \cref{eq:three_body_ho_basis} is extended to
\begin{multline}
  \ket{\mathcal{N}_\text{3N}N n  \,  \alpha_\text{3N}\alpha \, \mathcal{J}_\text{tot} \mathcal{M}_{\mathcal{J}_\text{tot}}} \equiv\\
  \ket{\mathcal{N}_\text{3N} N n  \{\mathcal{L}_\text{3N}[(L S) J (\ell \tfrac{1}{2})j] \mathcal{J} \} \mathcal{J}_\text{tot} \mathcal{M}_{\mathcal{J}_\text{tot}} (T \tfrac{1}{2}) \mathcal{T} \mathcal{M}_\mathcal{T}},
\end{multline}
where the subscript $\text{3N}$ denotes quantities related to the three-body center of mass.
This state corresponds to the coordinate representation in the left part of \cref{fig:coordinate_systems}.

The matrix elements of the Sachs operator in the three-body Jacobi basis can be obtained by carrying out the following transformation:
\begin{widetext}
\begin{multline}
\langle \mathcal{N}_{\rm 3N}' N'n' \alpha_{\rm 3N}' \alpha' \mathcal{J}_{\rm tot}' \mathcal{T}'||| \mu^{\rm NLO, Sachs}_{{\rm 2b},12} ||| \mathcal{N}_{\rm 3N}Nn \alpha_{\rm 3N}\alpha  \mathcal{J}_{\rm tot} \mathcal{T} \rangle \\
  \begin{split}
  =& \sum_{N_{12}' \, N_\text{NN}' \, n_3' } \sum_{\alpha'_{12} \, \alpha'_3} \sum_{N_{12} \, N_\text{NN} \, n_3}  \sum_{\alpha_{12} \, \alpha_3} \braket{\mathcal{N}_\text{3N}'N' n'  \, \alpha'_\text{3N} \, \alpha'  \, \mathcal{J}_\text{tot} }{N_\text{NN}' N_{12}'  n_3' \, \alpha'_{12} \, \alpha'_3 \, \mathcal{J}_\text{tot} } \\
 & \times \langle  N'_{\rm NN}N'_{12}  n'_{3}\alpha'_{12}\alpha'_{3} \mathcal{J}'_{\rm tot}\mathcal{T}' |||\mu^{\rm NLO, Sachs}_{{\rm 2b},12} ||| N_{\rm NN} N_{12} n_{3}\alpha_{12}\alpha_{3} \mathcal{J}_{\rm tot}\mathcal{T} \rangle  \\
  & \times \braket{N_\text{NN}N_{12}  n_3 \, \alpha_{12} \, \alpha_3 \, \mathcal{J}_\text{tot}}{N_\text{3N} N n  \, \alpha_\text{3N} \, \alpha  \, \mathcal{J}_\text{tot} }.
  \end{split}\label{eq:two_body_jacobi}
\end{multline}
\end{widetext}
The remaining task is to determine the basis transformation brackets $\braket{N_\text{NN} N_{12}  n_3 \, \alpha_{12} \, \alpha_3 \, \mathcal{J}_\text{tot} }{\mathcal{N}_\text{3N}N n  \, \alpha_\text{3N} \, \alpha  \, \mathcal{J}_\text{tot}}$, as the matrix element with respect to the three-body basis is already given in \cref{eq:sachs_trhee_body_me}.
Again, since the intrinsic and center-of-mass motions are exactly decoupled, we are free to choose any $\mathcal{N}_{\rm 3N}$ and $\mathcal{L}_{\rm 3N}$.
The most convenient choice is $\mathcal{N}_{\rm 3N}=\mathcal{L}_{\rm 3N}=0$.
Then, the overlap is found to be

\begin{equation}
\begin{aligned}
\label{eq:3b_overlap}
\langle N_{\rm NN} N_{12}& n_{3}\alpha_{12}\alpha_{3} \mathcal{J}_{\rm tot} | 0 Nn\alpha\mathcal{J}_{\rm tot} \rangle
=
(-1)^{L_{\rm NN}-J_{\rm rc}+ \ell_{3} + 1/2 + j_{3} + j + \mathcal{J}}
\\ & \quad \times
\hat{J}_{\rm rc} \hat{j} \hat{j}_{3}\hat{l}
\sixj{J}{L_{\rm NN}}{J_{\rm rc}}{j_{3}}{\mathcal{J}}{j}
\sixj{L_{\rm NN}}{\ell_{3}}{\ell}{1/2}{j}{j_{3}}
\\ & \quad \times
\langle 00 n\ell |  N_{\rm NN}L_{\rm NN}n_{3}\ell_{3}\rangle_{d=2}
\\ & \quad  \times
\delta_{NN_{12}} \delta_{LL_{12}} \delta_{SS_{12}} \delta_{JJ_{12}}\delta_{TT_{12}} \delta_{\mathcal{J}\mathcal{J}_{\rm tot}}.
\end{aligned}
\end{equation}
Here, the Talmi-Moshinsky brackets are necessary to transform between the three- and two-body center-of-mass systems.
Also note that the Kronecker deltas indicate the relative two-body quantum numbers of the two different bases to be the same, which is expected as they essentially represent the same subsystem.
A detailed derivation of this result can be found in \cref{sec:two_body_center_of_mass_dependent_operator}.
Note that the object in the left-hand side in Eq.~\eqref{eq:two_body_jacobi} takes the required form in Eq.~\eqref{eq:mu2_with_3body_ho} as we set $\mathcal{N}_{\rm 3N}=\mathcal{L}_{\rm 3N}=0$.

\section{Results}\label{sec:results}
In this section, we examine the magnetic form factors and the magnetic moments of the trinucleons using different nuclear interactions based on chiral EFT.
First, we present results for the magnetic form factors by evaluating the current operator in terms of the matrix elements presented in~\cref{subsec:current_operator_expansion} with corresponding partial-wave expanded wave functions, which are obtained by solving the three-body Faddeev equations.
Magnetic moments are calculated as the zero-momentum-transfer limits of these form factors.
We furthermore show results for the magnetic moments obtained from the expanded magnetic dipole operator expressions discussed in Sec.~\ref{subsec:dipole_operator_expansion}, based on NCSM wave functions.
Finally, we compare Faddeev and NCSM calculations against each other to benchmark our results for the magnetic dipole operator.

We use the non-local chiral NN interactions by Entem, Machleidt, and Nosyk (EMN)~\cite{EntemMachleidtNosyk2020} from LO to N$^3$LO with cutoffs $\Lambda = 420$, $450$, and $500$~MeV.
These are supplemented with 3N interactions at the same orders, with 3N low-energy constants (LECs) $c_D$ and $c_E$ determined by fits to the triton binding energy and nuclear matter saturation properties, and with a nonlocal three-body regulator with cutoff $\Lambda_\text{3N}$ identical to the NN cutoff~\cite{drischler2019}.
A systematic study of the dependence of the magnetic observables on the three-body (LECs) $c_D$ and $c_E$ is, however, not pursued, because the magnetic form factors turn out to be nearly independent of the 3N interaction, and the magnetic moments even less so~\cite{seutin2021}.
In addition, we also consider the Entem and Machleidt (EM)~\cite{entemmachleidt2003} interaction at N$^3$LO, with a cutoff $\Lambda = 500$~MeV.

The availability of the EMN potentials at each order of the expansion makes it possible to calculate theoretical uncertainty estimates of neglected higher-order terms based on the convergence pattern of observables.
We use a Bayesian model as outlined in~\cite{furnstahl2015a,melendez2017,melendez2019} to provide a statistical approach to calculate these uncertainties for the form-factor results.
This method determines a posterior distribution which captures all the information about the neglected higher-order terms, from which degree-of-belief (DoB) intervals are calculated.
For the evaluation, we employ a prior set $C_{0.25-10}$ with $\Lambda_\text{b}=650$~MeV, specified and publicly made available as a code in Ref.~\cite{melendez2017}.
A characteristic momentum scale of $p =2/3Q$ is used to calculate the $68\%$ and $95\%$ DoBs of the form factors.
Overall, the characteristic scale for momentum $Q$ transferred to a nucleus with mass number $A$ is set by $(A-1)/A \, Q$ according to Ref.~\cite{phillips2016}.

The trinucleon magnetic form factors and dipole moments are computed as
\begin{align}\label{eq:form_factor_ev}
  F_\text{M}(Q) = -\frac{2 m_N}{Q} \mel{\Psi^{\mathcal{J},\mathcal{T}}_{\mathcal{M}_\mathcal{J}+1, \mathcal{M}_\mathcal{T}}({\rm F})}{\, j_{+}(Q) \,}{\Psi^{\mathcal{J},\mathcal{T}}_{\mathcal{M}_\mathcal{J}, \mathcal{M}_\mathcal{T}}({\rm F})},
\end{align}
and
\begin{align}\label{eq:dipole_moment_ev}
  \mu_\text{t/h} = \mel{\Psi^{\mathcal{J},\mathcal{T}}_{\mathcal{M}_\mathcal{J}, \mathcal{M}_\mathcal{T}} ({\rm NCSM})}{\, \mu_{z} \,}{\Psi^{\mathcal{J},\mathcal{T}}_{\mathcal{M}_\mathcal{J}, \mathcal{M}_\mathcal{T}} ({\rm NCSM})},
\end{align}
respectively.
Here, $|\Psi^{\mathcal{J},\mathcal{T}}_{\mathcal{M}_\mathcal{J}}, ({\rm F})\rangle$ and $|\Psi^{\mathcal{J},\mathcal{T}}_{\mathcal{M}_\mathcal{J}}, ({\rm NCSM})\rangle$ are the Faddeev and NCSM wave functions, respectively, with total three-body spin and isospin specified by $\mathcal{J} = \mathcal{T} = 1/2$ and maximally projected total angular momentum states, i.e., $\mathcal{M}_\mathcal{J}=\pm \mathcal{J}$.
The isospin projection $\mathcal{M}_\mathcal{T} = {-}1/2$ or $\mathcal{M}_\mathcal{T} = 1/2$ determines whether the wave function represents a triton or a helion, respectively.

\begin{figure*}[t!]
\centering
  \includegraphics[width=0.8\textwidth]{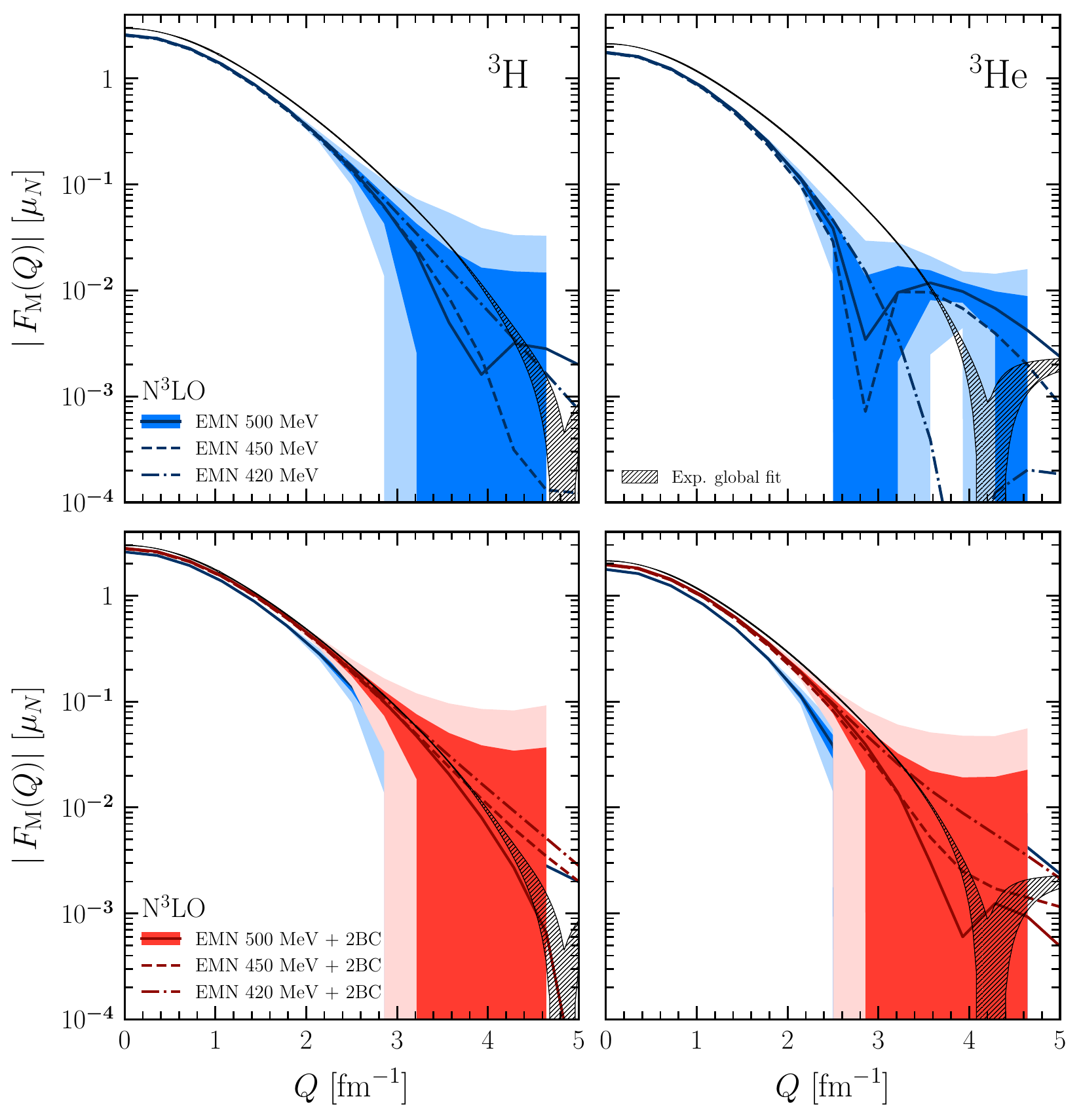}
  \caption{Triton (left column) and helion (right column) magnetic form factors, in units of $\mu_N$, as a function of the momentum transfer, in units of fm$^{-1}$. The hatched band represents a parametrization of the elastic scattering data~\cite{amroun1994}. The top row shows the result with the one-body current operator only, while the bottom row includes the 2BC contributions. The solid, dashed, and dashed-dotted lines represent the results for interactions at N$^3$LO with cutoff $\Lambda= 500$, $450$, and $420$ MeV, respectively. Light and dark shaded bands represent the $95\%$ and $68\%$ DoBs. }
  \label{fig:two_body_form_factor}
\end{figure*}
%

In order to perform Faddeev calculations, we truncate the basis by choosing a maximal value for the relative total two-body angular momentum $J$.
Our calculations include partial waves up to $J \leq 6$, which generates $42$ distinct combinations of one- and two-body quantum numbers.
This truncation proves to be sufficient to obtain converged results with respect to the basis states, so that any variation observed in the results are attributed to the interactions.

\subsection{Magnetic form factor}
We use the EMN interactions to calculate the magnetic form factors of the trinucleons with LO and NLO current operators.
\Cref{fig:two_body_form_factor} shows the triton (left column) and the helion (right column) magnetic form factors, in units of $\mu_N$, for calculations with the one-body current operator only (top row) and including 2BC corrections at NLO (bottom row) as a function of the momentum transfer $Q$, in units of fm$^{-1}$, and are compared to experimental results which are summarized by the hatched band~\cite{amroun1994}.
Results are given for cutoffs $\Lambda = 420$, $450$, and $500$~MeV at N$^3$LO by the dashed-dotted, dashed, and solid lines, respectively, together with the $68\%$ (light band) and $95\%$ (dark band) DoBs for the $500$~MeV result.
These bands terminate around $Q\sim4.6$ fm$^{-1}$, because the Bayesian method has a limited range of validity.
Lower orders of the interaction are used to calculate the truncation uncertainty, displayed by the colored bands, but not explicitly shown otherwise.
\begin{figure*}[t!]
\centering
  \includegraphics[width=0.8\textwidth]{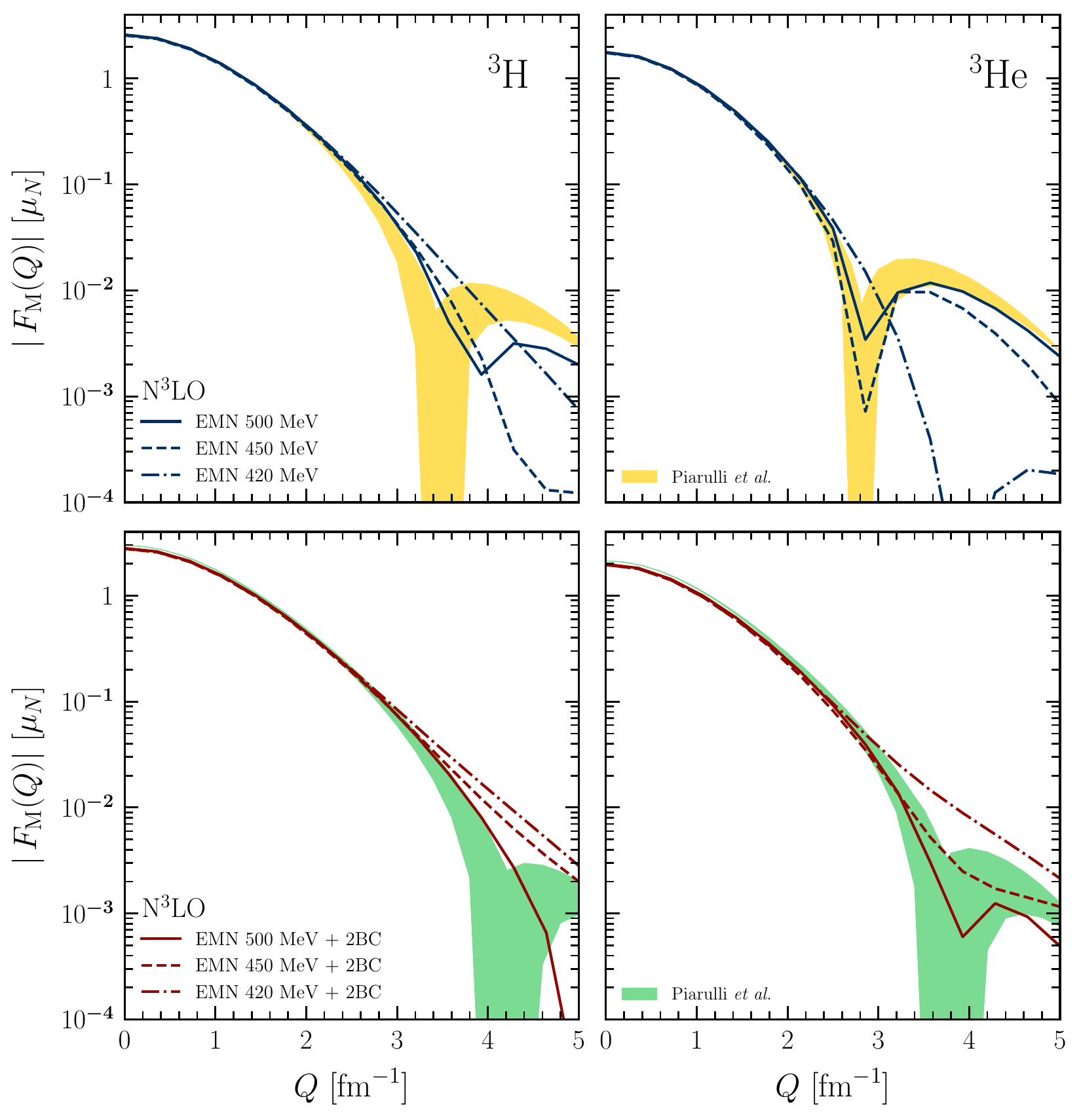}
  \caption{Same as \cref{fig:two_body_form_factor}, but compared to results from Ref.~\cite{piarulli2013} and without DoB intervals. The green and yellow bands correspond to a variation of the cutoff scale from $\Lambda = 500$ MeV to $\Lambda = 600$ MeV. }
  \label{fig:two_body_form_factor_comparison}
\end{figure*}

The magnetic form-factor results for $^3$He obtained with only the one-body current disagree with experiment for all cutoffs over the entire momentum-transfer region and underestimate the data until the minimum, even when the truncation uncertainty is considered.
For $^3$H, the results disagree below $3.8$ fm$^{-1}$ when the truncation uncertainty is taken into account, yet fall within the $95\%$ DoB interval at larger momentum transfer.
Although the truncation uncertainty clearly increases as $Q$ grows, note that the logarithmic scale overemphasizes the uncertainty at large $Q$ in the figures.
As expected, the cutoff dependence increases as well with momentum transfer, and the bands for the different cutoff values separate around $Q \sim 3$ fm$^{-1}$; nevertheless they remain within the $68\%$ DoB intervals.
At $Q = 0$, the results for $\Lambda=500$ MeV, $F_\text{M, t}(0)=2.583$ $\mu_N$ and $F_\text{M, h}(0)=-1.767$ $\mu_N$, deviate from the experimental values for the triton ($2.9789624659(59)$ $\mu_N$) and helion ($-2.127625307(25)$ $\mu_N$~\cite{codata2018}) magnetic moments, and within the low-momentum transfer regime an approximate constant offset from experiment is found.
In the following, we will examine the form factor normalization, i.e., the magnetic moment, in more detail.
It is well known that higher-order two-body current operator corrections are sizable at $Q = 0$~\cite{schiavilla1998,piarulli2013}, and thus they will impact the offset observed at low momentum transfers.

The bottom row of \cref{fig:two_body_form_factor} shows results with 2BC corrections included.
Values at low momentum transfer are shifted up for both nuclei, but still disagree with experiment.
Note that the DoB intervals of the one-body and two-body results do not overlap at low momentum transfers, which is a consequence of the inconsistent inclusion of current operators compared to the order of the interaction.
At higher $Q$, the minimum is shifted to higher momentum transfers, so that the central value of the helion band is slightly too low and the central value of the triton band reproduces the minimum, confirming once again that 2BCs provide essential corrections to the one-body current operator.
Because the chiral truncation uncertainty is expected to grow for increasing momentum transfers, the goal of exactly reproducing the minimum is too strong.
We observe that within the truncation uncertainty the higher $Q$ region fully overlaps with experimental data for both nuclei.
The cutoff variation is slightly reduced with respect to the one-body result and falls well within the $68\%$ DoBs.
Stronger conclusions can only be made by including higher-order operators in the calculation.

In \cref{fig:two_body_form_factor_comparison}, we compare our results to calculations from Piarulli {\it et al.}~\cite{piarulli2013}, which are given by the yellow and green bands.
The upper row displays the comparison of the one-body current operator (yellow band), while the bottom row shows results with 2BCs included (green band).
Their results are obtained by calculating the expectation value of the operators with wave functions generated by the hyperspherical harmonics framework.
The bands in this case represent the variation of the cutoff from $\Lambda = 500$ MeV to $\Lambda = 600$ MeV of the employed chiral interaction and therefore have a different interpretation compared to our bands.

Our one-body current results for $^3$He with cutoffs $\Lambda=450$ and $500$ MeV fall within the yellow band over the entire momentum range, while our results for $^3$H overestimate the minimum and the high momentum region.
Considering that the operator is identical, the differences could only be explained by the use of different chiral interactions.
This shows that apart from the corrections to the operator, also the interaction strongly influences the high momentum transfer region.

The green bands in the bottom row include 2BC operators up to N$^3$LO, whereas our results only include the leading NLO contributions.
At zero and low momentum transfers a small difference is observed, we will clarify its origin when discussing the magnetic moment below.
For the minimum and the high momentum region, the observation is similar to the one-body comparison.
Assuming that the truncation uncertainty for the results from Ref.~\cite{piarulli2013} would be comparable to our findings, all results at high momentum transfers would agree and even be consistent with experiment.

\begin{figure}[t]
\centering
  \includegraphics{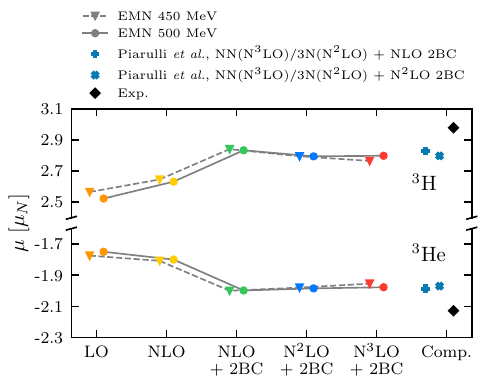}
  \caption{Triton and helion magnetic moments in units of $\mu_N$ as a function of increasing order of the chiral expansion for the EMN NN + 3N interaction at cutoffs $450$ MeV (\emph{downward triangles} and \emph{dashed lines}) and $500$ MeV (\emph{circles} and \emph{solid lines}). Results from calculations with LO, NLO, NLO + NLO 2BC, \ntwolo + NLO 2BC, and \nthreelo + NLO 2BC interactions are shown in orange, yellow, green, blue, and red, respectively. Blue plus and cross symbols show the magnetic moment including NLO and N$^2$LO current operator corrections, respectively, from Ref.~\cite{piarulli2013}. The black diamonds represent the experimental values for both nuclei~\cite{codata2018}. }
  \label{fig:form_factor_normalization_convergence}
\end{figure}

In order to systematically study the low momentum transfer region, we display the trinucleon magnetic moments obtained from $F_\textnormal{M}(0)$ in \cref{fig:form_factor_normalization_convergence} as a function of increasing chiral order, for cutoffs $450$ MeV (indicated by downward triangles and connected by dashed lines) and $500$ MeV (indicated by circles and solid lines).
They are compared to the experimental values of the triton and the helion, as well as to results from Piarulli~\textit{et al.}~\cite{piarulli2013} (blue pluses and crosses), which include NLO 2BC corrections and the relativistic one-body correction to the magnetic moment at N$^2$LO.

The magnetic moments can be understood as one-nucleon hole with respect to $^{4}$He, and the single-particle limit for triton (helion) of 2.793 (-1.913) $\mu_{N}$ is reasonably close to the computed result.
We observe that increasing the order of the chiral interaction used for the bound-state calculation has almost no effect on the magnetic moment of the trinucleons.
However, adding NLO 2BC corrections (shown as ``NLO + NLO 2BC'' in the figure) changes the values by ${\sim}10\%$ and improves agreement with experiment.
Our final result, labeled ``N$^3$LO + NLO 2BC,'' agrees well with both results from Piarulli~\textit{et al.}, which implies that the relativistic correction to the one-body operator is very small.
On the other hand, comparing the result to the experimental values suggests that important corrections to the operator are still missing to explain the remaining $5-7\%$ discrepancy.
Two-body corrections to the current operator at N$^3$LO introduce new LECs that have to be fixed before predictions can be made~\cite{pastore2009,koelling2009}.
Different strategies exist for this procedure, and commonly a combination of observables is chosen which includes the isoscalar $\mu_S$ and the isovector combination $\mu_V$ of the trinucleon magnetic moments to constrain the new LECs~\cite{pastore2009,piarulli2013}.\footnotemark
\footnotetext{The isoscalar ($\mu_S$) and isovector ($\mu_V$) combinations of the trinucleon magnetic moments are defined by $\mu_S \equiv \mu_\textnormal{t} + \mu_\textnormal{h}$ and $\mu_V \equiv \mu_\textnormal{t} - \mu_\textnormal{h}$.}
As a result, the experimental magnetic moments of the trinucleons are reproduced exactly if these higher-order corrections to the operator are taken into account.
Therefore, tests of higher order 2BC require finite momentum transfer or nuclei beyond $A=3$.

\subsection{Test of magnetic dipole operator}
\begin{figure*}[t]
\centering
    \includegraphics[width=\textwidth]{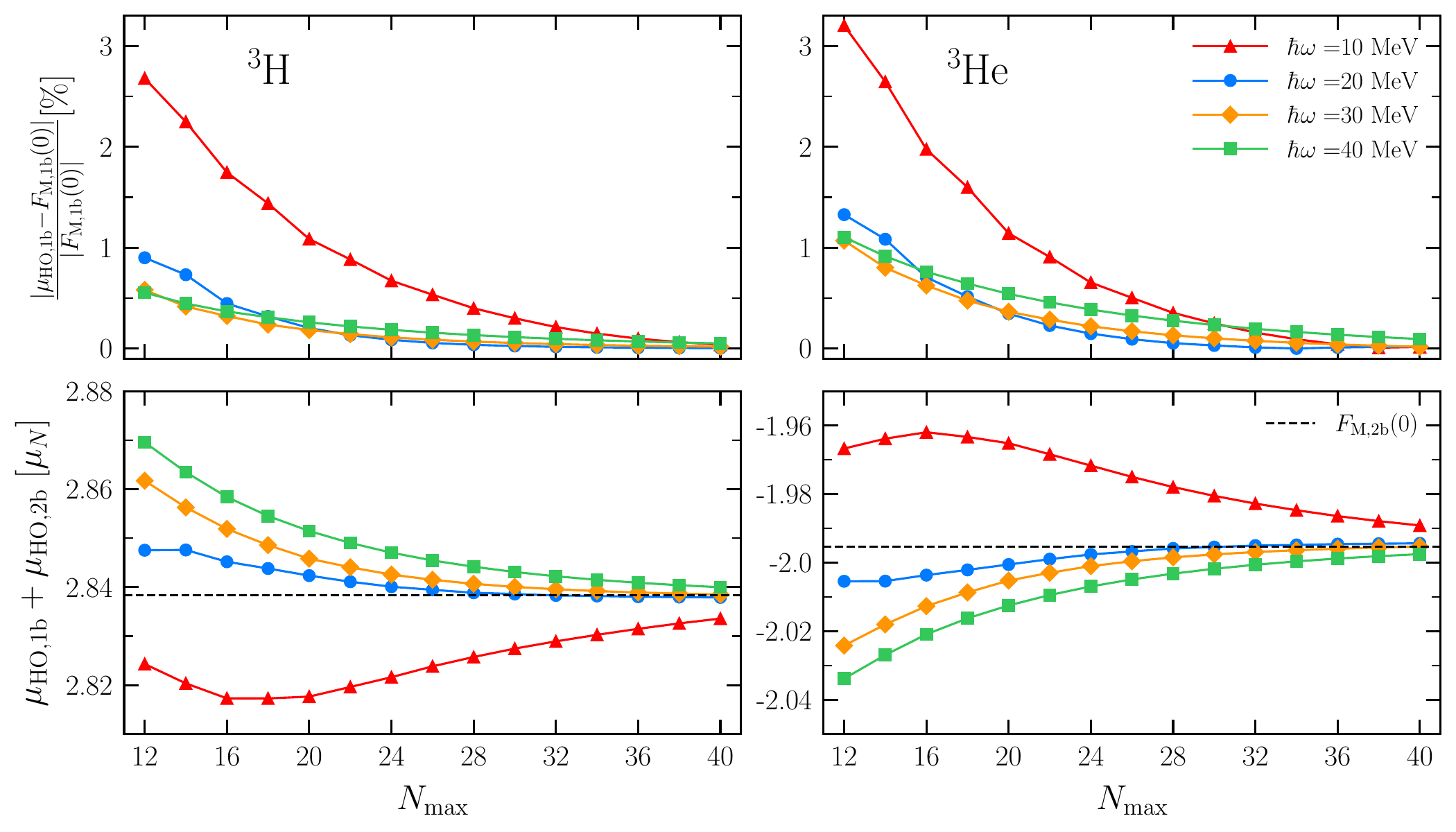}
  \caption{Convergence of the triton (left column) and helion (right column) magnetic moments as a function of $N_\text{max}$ for four different HO frequencies $\hbar\omega = 10, 20, 30,$ and $40$ MeV obtained with the EM NN interaction with cutoff $\Lambda = 500$~MeV.
  The upper row shows the relative deviation, in $\%$, of the magnetic moment calculated from the magnetic dipole operator compared to the form factor normalization, while the lower row shows the convergence of the magnetic moment including the NLO contributions, consisting of the intrinsic and Sachs components, towards the NLO form factor result (dashed line), in $\mu_N$.}
  \label{fig:lo_magnetic_moment_deviation}
\end{figure*}

In this section we present the trinucleon magnetic moments obtained from the magnetic dipole operator with the Jacobi NCSM, as discussed in~\cref{subsec:dipole_operator_expansion}.
The three-body NCSM calculations are done with \texttt{NuHamil} code~\cite{Miyagi2023}.
We examine their convergence behavior and benchmark them to the magnetic moments obtained from the form factors in momentum space.
Because our main goal is to benchmark the magnetic dipole operator matrix elements, we only consider the EM N$^3$LO interaction with cutoff $\Lambda = 500$~MeV, without 3N interactions.

\Cref{fig:lo_magnetic_moment_deviation} shows the convergence of the triton (left) and helion (right) magnetic moments as a function of $N_\text{max}$ for four different HO frequencies $\hbar\omega=10$, $20$, $30$, and $40$~MeV.
The top row displays the one-body magnetic dipole operator results relative to the one-body magnetic form factor normalization, while the bottom row shows the NLO-corrected result, represented by the label ``$\mu_\text{HO, 1b} + \mu_\text{HO, 2b}$'', with the corresponding form-factor normalization in absolute terms.
The two-body contribution consists of the intrinsic and Sachs terms, i.e., $\mu_\text{HO, 2b} = \mu^\text{NLO, intrinsic}_{2b} + \mu^\text{NLO, Sachs}_{2b}$, where the majority of the correction is contributed by the intrinsic component.

\begin{table}[t!]%
  \caption{Triton and helion magnetic moments and their (cumulative) contributions, in $\mu_N$, from the form factor normalization, the magnetic dipole operator, as well as experimental values~\cite{codata2018}. The NCSM wave function is computed at $N_{\rm max}=40$ and $\hbar\omega=20$ MeV.}
  \centering
  \begin{tabularx}{0.85\columnwidth}{l @{\hskip 5mm} ll c @{\hskip 5mm} ll}
    \toprule
   &  \multicolumn{2}{c}{$\mu_\textnormal{t}$ $[\mu_N]$} && \multicolumn{2}{c}{$\mu_\textnormal{h}$ $[\mu_N]$} \\
  \cmidrule{2-3}\cmidrule{5-6}
  & \multicolumn{1}{c}{$F_\textnormal{M}(0)$} & \multicolumn{1}{c}{$\hat{\boldsymbol{\mu}}$} && \multicolumn{1}{c}{$F_\textnormal{M}(0)$} & \multicolumn{1}{c}{$\hat{\boldsymbol{\mu}}$}  \\
  \midrule
  LO                     & $2.622    $ & $2.622    $ && $-1.783   $ & $-1.783   $ \\
  NLO                    & $2.838    $ & $2.837    $ && $-1.995   $ & $-1.994   $ \\
  \hspace{5mm} intrinsic & -           & $0.195    $ && -           & $-0.191   $ \\
  \hspace{5mm} Sachs     & -           & $0.021    $ && -           & $-0.021   $ \\
  \midrule
  Exp.                   & \multicolumn{2}{c}{$2.979$} && \multicolumn{2}{c}{$-2.128$} \\
  \bottomrule
  \end{tabularx}\label{tab:trinucleon_comparison}
\end{table}

Below we present numerical values and discuss the impact of both contributions in more detail.
The results show a systematic convergence towards the desired results obtained from the form factor normalizations.
Remarkably, very high values of $N_\text{max}$ are required in order to obtain converged results.
This pattern follows the slow convergence of the three-body energy, which is a consequence of describing a loosely bound system in a HO basis.

\Cref{tab:trinucleon_comparison} displays the magnetic moments obtained from both methods.
The left column contains the results for the triton magnetic moment $\mu_\text{t}$ from the form-factor normalization and the magnetic dipole operator and the right column gives the same results for the helion $\mu_\text{h}$.
The first row presents results from calculations with the one-body operator only (``LO''), while the second row shows results with the NLO 2BCs included (``NLO'').
Contributions from the latter to the magnetic moment are shown separately by the rows indicated with ``intrinsic'' and ``Sachs.''
As noted in~\cref{subsec:dipole_operator_expansion}, this separation cannot be made for the form factor calculation, hence only the total values can be compared.
At the bottom, the experimental results for both nuclei are given.
The effect on the ground-state magnetic moment of the $\hat{\boldsymbol{\mu}}^\textnormal{NLO, intrinsic}_\textnormal{2b}$ operator accounts for the bulk of the correction and its influence amounts to around $10\%$, while the Sachs operator, which requires much more resources to calculate, has a minor effect of $0.5$-$1$\%.
Total results from both methods agree with each other within $\ll 1\%$.

The excellent agreement between the magnetic moments of the triton and helion obtained from both methods gives strong confidence that the partial-wave decomposition of the LO and in particular the NLO magnetic dipole operator has been carried out correctly.
After a transformation to single-particle coordinates (given in Appendix~\ref{sec:TM_trans}), the matrix elements presented in this work can be used in many-body basis-expansion frameworks that are capable of calculating observables for heavier nuclei.
For example, a recent NCSM calculation of the magnetic moment and a magnetic transition of $^6$Li based on the developments presented in this work showed that the NLO corrections to the magnetic dipole operator are essential and improve the agreement with experiment~\cite{gayer2020}.

\section{Summary and conclusions \label{sec:summary}}

In this paper, we studied the nuclear magnetic dipole operator obtained from chiral EFT current operators with a particular focus on the two-body NLO contribution.
We discussed the general connection between the current operator and the magnetic dipole operator, and presented the coordinate space expressions for the NLO magnetic dipole operator.
The magnetic dipole operator from 2BCs can be split into two terms: the intrinsic and Sachs terms, where the Sachs term depends explicitly on the two-body center-of-mass, in addition to the relative coordinate.
We derived in detail the partial-wave decomposed matrix elements of the operators in the corresponding single-particle, two-body, and three-body bases.
For the current operator we employed momentum-space basis states, to easily accommodate for the momentum dependence of the operator, while the magnetic dipole operator was evaluated with respect to HO basis states through Eqs.~\eqref{eq:intrinsic_two_body_expansion} and~\eqref{eq:sachs_two_body_me}.
The decomposition of the latter has been checked by calculating the trinucleon magnetic moments and benchmarking them to form factor normalization results.

We provided results for the trinucleon magnetic form factors based on the EMN potentials, which allowed us to also estimate the uncertainty arising from truncating the chiral expansion.
This uncertainty estimate, which by default grows for increasing momentum transfer, indicates that the precise reproduction of the minimum is not an important discriminator for EFTs.
Our results agree well with values previously obtained in the literature using different chiral interactions and are consistent with experiment if the uncertainty in the EFT truncation is taken into account.
In addition, our results for the magnetic dipole operator show a very good convergence and agree well with the form factor normalization results, demonstrating that the coordinate-space expression and the partial-wave decomposition of the dipole operator are correct.

This work establishes a starting point for many-body expansion methods to incorporate the HO partial-wave decomposed NLO dipole matrix elements for calculations of electromagnetic observables.
Such studies could validate already obtained results for $A\leq 9$ systems~\cite{pastore2009,gayer2020} and will extend the {\it ab initio} analysis of NLO corrections to magnetic observables to medium-mass nuclei based on chiral EFT interactions and consistent current operators.
As a next step, the partial-wave decomposition of the magnetic dipole operator could be pushed to higher orders in the chiral expansion.
This will test the chiral expansion and reduce the truncation uncertainty.

\begin{acknowledgements}

We thank K. Wendt for useful discussions and checks.
The three-body NCSM calculations are done with \texttt{NuHamil} code~\cite{Miyagi2023}.
This work was supported in part by the BMBF Contract No.~05P21RDFNB, by the Deutsche Forschungsgemeinschaft (DFG, German Research Foundation) -- Project-ID 279384907 -- SFB 1245, the Cluster of Excellence ``Precision Physics, Fundamental Interactions, and Structure of Matter'' (PRISMA$^+$ EXC 2118/1, Project ID 39083149), by the European Research Council (ERC) under the European Union's Horizon 2020 research and innovation programme (Grant Agreement No.~101020842), by the Max Planck Society, and by the National Science Foundation under Grant No. PHY--2044632. This material is based upon work supported by the U.S. Department of Energy, Office of Science, Office of Nuclear Physics, under the FRIB Theory Alliance, award DE-SC0013617.

\end{acknowledgements}

\appendix

\section{Three-body Jacobi coordinate\label{sec:Jacobi_coodinate}}
For our Faddeev calculations, the Jacobi momenta are defined as
\begin{equation}
\left(
\begin{array}{c}
\v{K}_{\rm cm} \\
\v{p}\\
\v{q}
\end{array}
\right) =
\left(
\begin{array}{ccc}
1 & 1 & 1 \\
\frac{1}{2} & -\frac{1}{2} & 0 \\
-\frac{1}{3} & -\frac{1}{3} & \frac{2}{3}
\end{array}
\right)
\left(
\begin{array}{c}
\v{k}_{1} \\
\v{k}_{2} \\
\v{k}_{3}
\end{array}
\right),
\end{equation}
with the single-particle momenta $\v{k}_{i}$, $i=1,2,3$.
The corresponding conjugate coordinates are given by
\begin{equation}
\left(
\begin{array}{c}
\v{R}_{\rm cm} \\
\v{r}_{p}\\
\v{r}_{q}
\end{array}
\right) =
\left(
\begin{array}{ccc}
\frac{1}{3} & \frac{1}{3} & \frac{1}{3} \\
1 & -1 & 0 \\
-\frac{1}{2} & -\frac{1}{2} & 1
\end{array}
\right)
\left(
\begin{array}{c}
\v{r}_{1} \\
\v{r}_{2} \\
\v{r}_{3}
\end{array}
\right),
\end{equation}
with the single-particle coordinates $\v{r}_{i}$, $i=1,2,3$.
For our NCSM calculations, we use a symmetric choice as the Talmi-Moshinsky bracket~\cite{talmi1952,moshinsky1959} is defined with them.
The momenta and coordinates are then defined as
\begin{equation}
\left(
\begin{array}{c}
\boldsymbol{\pi}_{0} \\
\boldsymbol{\pi}_{1}\\
\boldsymbol{\pi}_{2}
\end{array}
\right) =
\left(
\begin{array}{ccc}
\sqrt{\frac{1}{3}} & \sqrt{\frac{1}{3}} & \sqrt{\frac{1}{3}} \\
\sqrt{\frac{1}{2}} & -\sqrt{\frac{1}{2}} & 0 \\
\sqrt{\frac{1}{6}} & \sqrt{\frac{1}{6}} & -\sqrt{\frac{2}{3}}
\end{array}
\right)
\left(
\begin{array}{c}
\v{k}_{1} \\
\v{k}_{2} \\
\v{k}_{3}
\end{array}
\right),
\end{equation}
and
\begin{equation}
\left(
\begin{array}{c}
\boldsymbol{\xi}_{0} \\
\boldsymbol{\xi}_{1}\\
\boldsymbol{\xi}_{2}
\end{array}
\right) =
\left(
\begin{array}{ccc}
\sqrt{\frac{1}{3}} & \sqrt{\frac{1}{3}} & \sqrt{\frac{1}{3}} \\
\sqrt{\frac{1}{2}} & -\sqrt{\frac{1}{2}} & 0 \\
\sqrt{\frac{1}{6}} & \sqrt{\frac{1}{6}} & -\sqrt{\frac{2}{3}}
\end{array}
\right)
\left(
\begin{array}{c}
\v{r}_{1} \\
\v{r}_{2} \\
\v{r}_{3}
\end{array}
\right).
\end{equation}

\section{Three-body overlap for Sachs term evaluation}
\label{sec:two_body_center_of_mass_dependent_operator}
Here we give a derivation of Eq.~\eqref{eq:3b_overlap}.
To this end, we consider the following recoupling so that we can factor out the orbital part:
\begin{equation}
\begin{aligned}
|N_{\rm NN} & N_{12} n_{3} \alpha_{12} \alpha_{3} \mathcal{J}_{\rm tot} \rangle
\\ & =
(-1)^{j_{3}+\mathcal{J}_{\rm tot}-J_{\rm rc}+L_{\rm NN}+\ell_{3}+1/2}
\\ & \quad \times
\sum_{\Lambda} (-1)^{\Lambda}
\hat{J_{\rm rc}}\hat{\Lambda}
\sixj{J_{12}}{L_{\rm NN}}{J_{\rm rc}}{j_{3}}{\mathcal{J}_{\rm tot}}{\Lambda}
\\ & \quad \times
\sum_{\lambda} \hat{\lambda}\hat{j_{3}}
\sixj{L_{\rm NN}}{\ell_{3}}{\lambda}{1/2}{\Lambda}{j_{3}}
\\ & \quad \times
  | N_{\rm NN} N_{12} n_3 \{ J_{12} \, [(L_{\rm NN}\ell_{3})\lambda \, \tfrac{1}{2}]\Lambda\} \mathcal{J}_\text{tot} \rangle.
\end{aligned}
\label{eq:3b_recoupling1}
\end{equation}
To factorize the three-body center-of-mass part, one can use the coordinate transformation
\begin{equation}
\left(
\begin{array}{c}
\boldsymbol{\xi}_{0}\\
\boldsymbol{\xi}_{2}
\end{array}
\right) = \left(
\begin{array}{cc}
\sqrt{\frac{2}{3}} & \sqrt{\frac{1}{3}} \\
\sqrt{\frac{1}{3}} & -\sqrt{\frac{2}{3}}
\end{array}
\right)
 \left(
 \begin{array}{c}
 \sqrt{\frac{1}{2}}(\v{r}_{1}+\v{r}_{2})\\
\v{r}_{3}
\end{array}
 \right).
\end{equation}
The above coordinate transformation ensures the following transformation using the Talmi-Moshinsky bracket~\cite{talmi1952,moshinsky1959}, with the notation given in Ref.~\cite{Kamuntavicius2001}:
\begin{equation}
\begin{aligned}
|N_{\rm NN}& N_{12} n_{3} \{J_{12} [(L_{\rm NN}\ell_{3})\lambda \frac{1}{2}]\Lambda\} \mathcal{J}_{\rm tot} \rangle
\\ &= \sum_{\mathcal{N}_{\rm 3N}' \mathcal{L}_{\rm 3N}' n'\ell'}
|\mathcal{N}_{\rm 3N}' N_{12} n' \{J_{12} [(\mathcal{L}_{\rm 3N}\ell')\lambda \frac{1}{2}]\Lambda\} \mathcal{J}_{\rm tot} \rangle
\\ & \quad \times
\langle \mathcal{N}_{\rm 3N}' \mathcal{L}_{\rm 3N}', n'\ell':\lambda |
N_{\rm NN} L_{\rm NN}, n_{3}\ell_{3}:\lambda \rangle_{d=2}.
\end{aligned}
\label{eq:3b_TM}
\end{equation}
For the Jacobi three-body basis, the recoupling is given by
\begin{equation}
\begin{aligned}
|\mathcal{N}_{\rm 3N}' & N_{12} n' \{J_{12} [(\mathcal{L}_{\rm 3N}\ell')\lambda \frac{1}{2}]\Lambda\} \mathcal{J}_{\rm tot} \rangle
\\&=
\sum_{j'} (-1)^{\ell'+1/2+j'} \hat{\lambda}\hat{j'}
\sixj{\mathcal{L}_{\rm 3N}'}{\ell'}{\lambda}{1/2}{\Lambda}{j'}
\\ & \quad \times
\sum_{\mathcal{J}'} (-1)^{J_{12}+j'+\mathcal{J}'}
\hat{\mathcal{J}}'\hat{\Lambda} \sixj{J_{12}}{j'}{\mathcal{J}'}{\mathcal{L}_{\rm 3N}'}{\mathcal{J}_{\rm tot}}{\Lambda}
\\ & \quad \times
|\mathcal{N}_{\rm 3N}' N_{12}n' [\mathcal{L}_{\rm 3N}'(J_{12}j')\mathcal{J}']\mathcal{J}_{\rm tot} \rangle.
\end{aligned}
\label{eq:3b_recoupling2}
\end{equation}
Combining Eqs.~\eqref{eq:3b_recoupling1}, \eqref{eq:3b_TM}, and \eqref{eq:3b_recoupling2}, and using the diagonality of the states, we obtain
\begin{equation}
\begin{aligned}
\langle \mathcal{N}_{\rm 3N}& Nn \alpha_{\rm 3N} \alpha \mathcal{J}_{\rm tot} |
\mathcal{N}_{\rm 3N}' N_{12}n' [\mathcal{L}_{\rm 3N}'(J_{12}j')\mathcal{J}']\mathcal{J}_{\rm tot} \rangle
\\ & =
\delta_{\mathcal{N}_{\rm 3N}\mathcal{N}_{\rm 3N}'}
\delta_{\mathcal{L}_{\rm 3N}\mathcal{L}_{\rm 3N}'}
\delta_{NN_{12}}
\delta_{LL_{12}}
\delta_{SS_{12}}
\delta_{JJ_{12}}
\delta_{nn'}
\delta_{\ell\ell'}
\delta_{jj'}
\delta_{\mathcal{J}\mathcal{J}'},
\end{aligned}
\end{equation}
and one can find the overlap:
\begin{equation}
\begin{aligned}
\langle  N_{\rm NN} & N_{12} n_{3}\alpha_{12}\alpha_{3} \mathcal{J}_{\rm tot} | \mathcal{N}_{\rm 3N} Nn\alpha_{\rm 3N}\alpha\mathcal{J}_{\rm tot} \rangle
 \\ & =
(-1)^{J_{12}+L_{12}-J_{\rm rc}+\ell_{3}+\ell+\mathcal{J}+\mathcal{J}_{\rm tot}}
\\ & \quad \times
\sum_{\Lambda\lambda}(-1)^{j_{3}+\Lambda}
\hat{J}_{\rm rc}\hat{\mathcal{J}} \hat{j}_{3} \hat{j}  \hat{\Lambda}^{2} \hat{\lambda}^{2}
\\ & \quad \times
\sixj{J_{12}}{L_{\rm NN}}{J_{\rm rc}}{j_{3}}{\mathcal{J}_{\rm tot}}{\Lambda}
\sixj{J_{12}}{j}{\mathcal{J}}{\mathcal{L}_{\rm 3N}}{\mathcal{J}_{\rm tot}}{\Lambda}
\\ & \quad \times
\sixj{L_{\rm NN}}{\ell_{3}}{\lambda}{1/2}{\Lambda}{j_{3}}
\sixj{\mathcal{L}_{\rm 3N}}{\ell}{\lambda}{1/2}{\Lambda}{j}
\\ & \quad \times
\langle \mathcal{N}_{\rm 3N}' \mathcal{L}_{\rm 3N}', n'\ell':\lambda |
N_{\rm NN} L_{\rm NN}, n_{3}\ell_{3}:\lambda \rangle_{d=2}
\\ & \quad \times
\delta_{NN_{12}}
\delta_{LL_{12}}
\delta_{SS_{12}}
\delta_{JJ_{12}}.
\end{aligned}
\label{eq:3b_overlap_general}
\end{equation}
Inserting $\mathcal{N}_{\rm 3N}=\mathcal{L}_{\rm 3N}=0$ into Eq.~\eqref{eq:3b_overlap_general} leads to Eq.~\eqref{eq:3b_overlap}.

\section{Transformation to single-particle basis \label{sec:TM_trans}}
For applications to medium-mass and heavier systems, current developments will need to be combined with basis-expansion methods such as coupled-cluster theory~\cite{Hagen2014} or the in-medium-similarity renormalization group~\cite{Hergert2016,Stroberg2019}.
Then, the matrix elements need to be expressed in terms of the single-particle coordinates rather than relative and center-of-mass coordinates.
They are related by Talmi-Moshinsky transformations, already mentioned in Appendex~\ref{sec:two_body_center_of_mass_dependent_operator}.

Our goal here is to show the transformation for
\begin{equation}
\langle n'_{1}\ell'_{1}j'_{1}, n'_{2}\ell'_{2}j'_{2}: J_{\rm tot}' || \mathcal{O}^{\lambda} || n_{1}\ell_{1}j_{1}, n_{2}\ell_{2}j_{2} : J_{\rm tot}\rangle.
\end{equation}
Here the state $|n_{1}\ell_{1}j_{1}, n_{2}\ell_{2}j_{2} :J_{\rm tot} \rangle$ is the antisymmetrized product of $\{n_{1}, \ell_{1}, j_{1}\}$ and $\{n_{2}, \ell_{2}, j_{2}\}$ states coupling to the total angular momentum $J_{\rm tot}$ in the proton-neutron ($p$-$n$) basis.
One can find the following transformation
\begin{equation}
\begin{aligned}
|n_{1}\ell_{1}j_{1}, & n_{2}\ell_{2}j_{2} :J_{\rm tot} \rangle =
\sum_{N_{12}N_{\rm NN}\alpha_{12}} |N_{\rm NN}N_{12}\alpha_{12} \rangle
\\ &\times
\langle N_{\rm NN}N_{12} \alpha_{12} | n_{1}\ell_{1}j_{1}, n_{2}\ell_{2}j_{2} :J_{\rm tot} \rangle,
\end{aligned}
\end{equation}
where the overlap is
\begin{equation}
\begin{aligned}
\langle N_{\rm NN}N_{12} & \alpha_{12} | n_{1}\ell_{1}j_{1}, n_{2}\ell_{2}j_{2} :J_{\rm tot} \rangle = (-1)^{L+L_{\rm NN}+S_{12}+J_{\rm tot}} f_{12}
\\ & \times
\hat{j}_{1}\hat{j}_{2}\hat{S}_{12}\hat{J}_{12}
\sum_{\Lambda} \hat{\Lambda}^{2} \ninej{\ell_{1}}{s_{1}}{j_{1}}{\ell_{2}}{s_{2}}{j_{2}}{\Lambda}{S_{12}}{J_{\rm tot}}
\\ & \times
\sixj{S_{12}}{L_{12}}{J_{12}}{L_{\rm NN}}{J_{\rm tot}}{\Lambda} \delta_{J_{\rm rc}J_{\rm tot}}
\\ & \times
\langle  N_{\rm NN} L_{\rm NN},N_{12} L_{12}: \Lambda |n_{1} \ell_{1}, n_{2} \ell_{2}: \Lambda \rangle_{d=1},
\end{aligned}
\end{equation}
with antisymmetrization factor
\begin{equation}
f_{12} = \left\{\!\!\!
\begin{array}{cc}
\sqrt{\frac{1}{2(1+\delta_{n_{1}n_{2}}\delta_{\ell_{1}\ell_{2}}\delta_{j_{1}j_{2}})}} [1+(-1)^{L_{12}+S_{12}}] & pp \textrm{ or } nn \\
1 & pn
\end{array}
\right. .
\end{equation}
Note that we have removed the isospin part from $\alpha_{12}$ as we work in the proton-neutron formalism.
Then, the matrix elements in the laboratory frame can be computed as
\begin{equation}
\begin{aligned}
\langle & n_{1}'\ell_{1}'j_{1}',  n_{2}'\ell_{2}'j_{2}' : J_{\rm tot}' || O^{\lambda} ||
n_{1}\ell_{1}j_{1}, n_{2}\ell_{2}j_{2} : J_{\rm tot} \rangle
\\ &= \sum_{N_{12}'N_{\rm NN}'\alpha_{12}'}
\sum_{N_{12}N_{\rm NN}\alpha_{12}}
\langle n_{1}'\ell_{1}'j_{1}', n_{2}'\ell_{2}'j_{2}' : J_{\rm tot}' |  N_{\rm NN}'N_{12}' \alpha_{12}' \rangle
\\ & \quad \times  \langle N_{\rm NN}'N_{12}'\alpha_{12}' || O^{\lambda} || N_{\rm NN}N_{12}\alpha_{12} \rangle
\\ & \quad \times
\langle N_{\rm NN}N_{12} \alpha_{12} | n_{1}\ell_{1}j_{1}, n_{2}\ell_{2}j_{2} : J_{\rm tot} \rangle.
\end{aligned}
\end{equation}
If the operator does not depend on the center-of-mass coordinate, the matrix element can be evaluated as
\begin{equation}
\begin{aligned}
\langle N_{\rm NN}' N_{12}'&\alpha_{12}' || O^{\lambda} || N_{\rm NN}N_{12}\alpha_{12} \rangle
= (-1)^{L_{\rm NN}+J_{\rm rc}+J'_{12}+\lambda}
\\ & \times
\hat{J}_{\rm rc}'\hat{J}_{\rm rc}
\sixj{J_{\rm rc}'}{J_{12}'}{L_{\rm NN}}{J_{12}}{J_{\rm rc}}{\lambda}
\langle N'_{12}\alpha_{\rm 2b}' || O^{\lambda} || N_{12}\alpha_{\rm 2b} \rangle
\\ & \times
\delta_{N_{\rm NN}N_{\rm NN}'} \delta_{L_{\rm NN}L_{\rm NN}'}.
\end{aligned}
\end{equation}

\bibliographystyle{apsrev4-2}
\bibliography{me_paper.bib}

\end{document}